\documentclass[10pt, a4paper]{article}
\usepackage{float}
\usepackage{subfig}
\usepackage{amsmath}
\usepackage{xcolor}
\usepackage{amssymb}
\usepackage{graphicx}
\usepackage{hyperref}
\usepackage{csquotes}
\usepackage{multirow}
\usepackage[left=.45in,right=.45in,top=.6in,bottom=.6in,headheight=14.5pt]{geometry}
\usepackage{multirow}
\usepackage{pdflscape}
\usepackage{appendix}
\usepackage[left=.45in,right=.45in,top=.6in,bottom=.6in,headheight=14.5pt]{geometry}
\usepackage{fmtcount} 
\usepackage{array,multirow}

\newcolumntype{C}[1]{>{\centering\arraybackslash}m{#1}}
\usepackage{epsfig}
\usepackage{float}
\usepackage{geometry}
\usepackage{amsmath}
\usepackage{cite}
\usepackage{ctable}

\newgeometry{vmargin={10mm,20mm},hmargin={19mm,19mm}}

\begin{document}

\title{Investigating Two-zero Textures of Inverse Neutrino Mass Matrix under the Lamp Post of LMA and LMA-D Solutions and Symmetry Realization}

\author{Labh Singh\thanks{sainilabh5@gmail.com}, Monal Kashav\thanks{monalkashav@gmail.com} and Surender Verma\thanks{s\_7verma@yahoo.co.in, Corresponding Author}}

\date{%
Department of Physics and Astronomical Science\\
Central University of Himachal Pradesh\\
Dharamshala, India 176215
}

\maketitle

\begin{abstract}

\noindent In this work we have investigated the phenomenological consequences of two-zero textures of \textit{inverse} neutrino mass matrix ($M_{\nu}^{-1}$) in light of the large mixing angle (LMA) and large mixing angle-\textit{dark} (LMA-D) solutions, later of which originates if neutrinos exhibit non-standard interactions with matter. Out of fifteen possibilities, only seven two-zero textures of $M_{\nu}^{-1}$ are found to be phenomenologically allowed under LMA and/or LMA-D descriptions. In particular, five textures are in consonance with both LMA and LMA-D solutions and are necessarily $CP$ violating while remaining two textures are found to be consistent with LMA solution only. The textures with vanishing (1, 1) element of $M_{\nu}^{-1}$ are, in general, disallowed. All the textures allowed under LMA and LMA-D solutions follow the same neutrino mass hierarchy. Furthermore, textures with vanishing (2, 3) element of  $M_{\nu}^{-1}$ are found to be either disallowed or are consistent with LMA description only.  We have, also, obtained the implication of the model for $0\nu\beta\beta$ decay amplitude $|M_{ee}|$. For most of the textures the calculated $3\sigma$ lower bound on $|M_{ee}|$ is $\mathcal{O}(10^{-2})$, which is within the sensitivity reach of $0\nu\beta\beta$ decay experiments. We have, also, proposed a flavor model based on discrete non-Abelian flavor group $A_4$ wherein such textures of $M_{\nu}^{-1}$ can be realized within Type-I seesaw setting.

\end{abstract}
\section{Introduction}
\noindent The experimental evidences accumulated in last two decades have convincingly established that not only neutrino have non-zero mass but its dynamical origin is beyond our current understanding of the standard model (SM). Despite the best efforts to precisely decipher the structure of neutrino mixing matrix, it still has certain unknowns, for example, $CP$ violating phases, octant of $\theta_{23}$ and neutrino mass hierarchy, to name a few. The theoretical frameworks to explain neutrino mass and its manifestations like neutrino oscillations have been developed assuming standard (charged (CC) and neutral current (NC)) interactions between neutrino and matter.  These frameworks culminated in large mixing angle (LMA) solution to the solar neutrino problem (SNP) and is well established by the neutrino oscillation experiments. In fact LMA has been independently confirmed as the solution to solar neutrino problem (SNP) in solar and KamLAND reactor experiments\cite{Petcov:2001sy}. It has been shown that for positive solar mass-squared difference, $\sin^2\theta_{12}$ cannot be greater than 0.5. However, at the subdominant level, there still have possibilities for additional contributions to neutrino oscillations such as non-standard interactions (NSI) of neutrino with matter fields\cite{Wolfenstein:1977ue,Proceedings:2019qno,Biggio:2009nt}. The future hyper-technological experiments will have the access to these unexplored regions. In presence of NSI, $\theta_{12}$ can be in the second octant. Thus, solar neutrino problem may have another degenerate solution in which $\sin^2\theta_{12}\approx0.7$. This solution is termed as large mixing angle-\textit{dark}  (LMA-D) solution. In general, the degeneracy between LMA and LMA-D solutions cannot be alleviated by oscillation experiments due to generalized mass hierarchy degeneracy in presence of NSI. However, a combined measurements from neutrino oscillation and scattering experiments may have imperative implication with regard to lifting of these degeneracies\cite{d1,d2,d3}.  Although non-standard interactions of neutrino with matter are severely constrained, the latest global fit, incorporating neutrino oscillation and COHERENT data, still allows LMA-D solution at $3\sigma$ confidence level\cite{cohe}. The only difference in LMA and LMA-D is in the octant of $\theta_{12}$, however, the solar mass-squared difference remains the same.

\noindent The progress in understanding the origin of neutrino mass centrally involves explaining the observed pattern of neutrino mixing which is encoded in the neutrino mass matrix obtained after electroweak symmetry breaking. Assuming neutrino to be Majorana particle the mass matrix contain nine free parameters \textit{viz.} three neutrino mass eigenvalues, three mixing angles and three CP violating phases. The two mass-squared differences and three mixing angles have been measured in neutrino oscillation experiments with high degree of precision. Seesaw mechanism is a natural and most effective way to explain the smallness of neutrino mass. Within Type-I seesaw\cite{Minkowski:1977sc,Mohapatra:1980yp} paradigm, the low energy effective neutrino mass matrix ($M_\nu$) is generated from Dirac neutrino mass matrix ($M_D$) and heavy right-handed Majorana neutrino mass matrix ($M_R$) using the relation: $M_\nu\approx M_DM_R^{-1}M_D^{T}$. The existence of near degeneracy in LMA and LMA-D solutions have imperative implications for models of neutrino mass and associated phenomenology. Recently, the LMA and LMA-D phenomenology of Majorana neutrino mass matrix has been studied assuming (i) zero textures of the neutrino mass matrix, $M_{\nu}$ \cite{Borgohain:2020now,Ghosh:2020fes,Zhao:2020cjm} (ii) in presence of one-sterile neutrino\cite{Deepthi:2019ljo}. Texture-zeros in the effective low energy Majorana neutrino mass matrix may have seesaw origin in which they can be realized from the zeros in $M_D$ and $M_{R}$\cite{Borah:2017azf}. In literature, there have been phenomenological studies with texture one-zero\cite{onezero1,onezero2,onezero3,onezero4} and two-zeros \cite{twozero1,twozero2,twozero3,twozero4,twozero5} while three-zeros or more, in neutrino mass matrix, are ruled out by current neutrino oscillation data\cite{msingh}. In Type-I seesaw, an interesting scenario may emerge if we work in $M_D$-diagonal basis. In this basis, the zero(s) in $M_\nu^{-1}$ is same as the the zero(s) in $M_R$ i.e. $M_\nu^{-1}\approx M_D^{-1}M_RM_D^{-1}$ \cite{Lavoura:2004tu}. In Ref. \cite{ot}, the authors have investigated the phenomenological consequences of one-zero texture in $M_\nu^{-1}$ with in the context of trimaximal mixing. The LMA phenomenology of two-zero textures of $M_{\nu}^{-1}$ has been investigated in \cite{Verma:2011kz}. It is to be noted that the texture zeros in $M_\nu$ and $M_\nu^{-1}$ are, in general, independent and may have distinguishing phenomenology. Motivated by the capabilities of the future neutrino oscillation experiments in resolving these subdominant effects\cite{Choubey:2019osj} and its important model building perspective, we investigate the phenomenological consequences of two-zero textures of $M_{\nu}^{-1}$. Also, we have proposed a flavor model based on non-Abelian discrete group $A_4$ and Type-I seesaw, where such zeros can be realized in the inverse neutrino mass matrix. There are fifteen possibilities to have two zeros in $M_{\nu}^{-1}$. We investigate the LMA and LMA-D phenomenology of these fifteen textures. 

\noindent The paper is organized as follows. In Sec. II, we
briefly introduce the formalism of two-zero texture of $M_{\nu}^{-1}$ and the details of the numerical analysis. Sec. III is devoted to the investigation and discussion of LMA/LMA-D phenomenology of the seven allowed textures of $M_{\nu}^{-1}$. A flavor model based on non-Abelian $A_4$ symmetry is discussed in Sec. IV. Finally, we brief our conclusions in Sec. V.
\section{Formalism of Two-zero Textures of Inverse Neutrino Mass matrix}
In charged lepton basis, the neutrino mass matrix is given by
\begin{equation}
    M_{\nu}=VM^{diag}_{\nu}V^{T},
    \label{1}
\end{equation}\
\noindent where $M^{diag}_{\nu}$= $diag(m_1,m_2,m_3)$ is the neutrino mass eigenvalue matrix, $V=U.P$ is complex unitary neutrino mixing matrix. The matrix $U$ is Pontecorvo-Maki-Nakagawa-Sakata (PMNS) matrix which in the PDG representation can be written as \cite{ParticleDataGroup:2020ssz}
\begin{equation}
U =\begin{pmatrix}
U_{11} & U_{12}  & U_{13} \\
U_{21} & U_{22}  & U_{23} \\
U_{31} & U_{32}  & U_{33}\\
\end{pmatrix}=
\begin{pmatrix}
c_{12}c_{13} & s_{12}c_{13}  & s_{13}e^{-i\delta} \\
-s_{12}c_{23}-c_{12}s_{23}s_{13}e^{i\delta} & c_{12}c_{23}-s_{12}s_{23}s_{13}e^{i\delta}  & s_{23}c_{13} \\
s_{12}s_{23}-c_{12}c_{23}s_{13}e^{i\delta} & -c_{12}s_{23}-s_{12}c_{23}s_{13}e^{i\delta}  & c_{23}c_{13} 
\end{pmatrix} ,
 \label{2}
\end{equation}\\
where $c_{ij}=cos\theta_{ij}$ and $s_{ij}=sin\theta_{ij}$ and $\delta$ is Dirac-type $CP$ violating phase. Also, $P$ is the phase matrix given by
\begin{equation*}
P = 
\begin{pmatrix}
1 & 0  & 0 \\
0 & e^{i\alpha}  & 0 \\
0 & 0  & e^{i(\beta+\delta)}\\
\end{pmatrix},
\end{equation*}\\
where $\alpha$, $\beta$ are Majorana-type $CP$ violating phases. The inverse neutrino mass matrix can be derived from Eqn. (\ref{1}) as
\begin{equation}
     M^{-1}_{\nu}=(VM^{diag}_{\nu}V^{T})^{-1}.
      \label{3}
\end{equation}
Using Eqns. (\ref{1}) and (\ref{2}), six independent elements of ($M^{-1}_{\nu}$) can be written as 
\begin{equation}
\left.
\begin{aligned}
(M^{-1}_{\nu})_{11}&=\dfrac{1}{m_{1}m_{2}m_{3}}\left[c^2_{13} e^{-2i\alpha}m_{3}(c^{2}_{12}e^{2i\alpha}m_{2}+m_{1}s^{2}_{12})+e^{-2i\beta}s^2_{13}m_{1}m_{2}\right], \quad \\
(M^{-1}_{\nu})_{12}&=\dfrac{1}{m_{1}m_{2}m_{3}}\left[e^{-i(2(\alpha+\beta)+\delta)}(c_{13}e^{2i\beta}m_{1}m_{3}s_{12}(c_{12}c_{23}e^{i\delta}-s_{12}s_{13}s_{23})
\right. \\ \left.
\right. & \left.
+c_{13}e^{2i\alpha}m_{2}(-c_{12}c_{23}e^{i(2\beta+\delta)}m_{3}s_{12})+(-c^{2}_{12}e^{2i\beta}m_{3}+m_{1})s_{13}s_{23})\right], \quad \\
(M^{-1}_{\nu})_{13}&=\dfrac{1}{m_{1}m_{2}m_{3}}\left[e^{-i(2(\alpha+\beta)+\delta)}(c_{13}c_{12}(e^{2i\alpha}m_{1}m_{2}-e^{2i\alpha}m_{3}(c^{2}_{12}e^{2i\alpha}m_{2}+m_{1}s^2_{12}))s_{13}
\right. \\ \left.
\right. & \left.
-c_{12}c_{13}e^{i(2\beta+\delta)}(m_{1}-e^{2i\alpha}m_{2})m_{3}s_{12}s_{23})\right],      \quad \\
(M^{-1}_{\nu})_{22}&=\dfrac{1}{m_{1}m_{2}m_{3}}\left[e^{-2i(\alpha+\delta)}m_1 m_3(c_{12}c_{23}e^{i\delta}-s_{12}s_{13s_{23}})^2 +m_{2}(c^2_{23}m_{3}s^2_{12}+2c_{12}c_{23}m_{3}s_{12}s_{13}s_{23}
\right. \\ \left.
\right. & \left.
+e^{-2i(\beta+\delta)}(c^2_{13}m_{1}+c^2_{12}e^{i\beta}m_{3}s^2_{13})s^2_{23})\right], \quad \\
(M^{-1}_{\nu})_{23}&=\dfrac{1}{m_{1}m_{2}m_{3}}\left[e^{-2i(\alpha+\beta)}m_{1}m_{3}(c_{23}s_{12}s_{13}+c_{12}e^{i\delta}s_{23})(c_{12}c_{23}e^{i\delta}-s_{12}s_{23}s_{13})
\right. \\ \left.
\right. & \left.
+e^{-2i(\beta+\delta)}m_{2}(c_{23}(c^2_{13}m_{1}-e^{2i(\beta+\delta)}m_{3}s^2_{12})+c^2_{12}c_{23}e^{i\beta}m_{3}s^2_{13}s_{23}
\right. \\ \left.
\right. & \left.
+c_{12}e^{i(2\beta+\delta)}m_{3}s_{12}s_{23}(c^2_{23}-s^2_{23}))\right], \quad  \\
(M^{-1}_{\nu})_{33}&=\dfrac{1}{m_{1}m_{2}m_{3}}\left[e^{-2i(\alpha+\beta+\delta)}c^2_{23}(c^2_{13}e^{2i\alpha}m_{1}m_{2}+e^{2i\beta}m_{3}(c^2_{12}e^{2i\alpha}m_{2}+m_{1}s^2_{12})s^2_{13})
\right. \\ \left.
\right. & \left.
 +e^{i(2\alpha+\delta)}2c_{12}c_{23}(m_{1}-e^{2i\alpha}m_{2})m_{3}s_{12}s_{13}s_{23}+e^{-2i\alpha}m_{3}(c^2_{12}m_{1}+e^{2i\alpha}m_{2}s^2_{12})s^2_{23}\right].
\end{aligned}
\right\}
 \label{4}
\end{equation}\\\

\noindent There are fifteen possible two-zero textures of $M^{-1}_{\nu}$ which we categorize in five classes \textit{viz.} class $A,B,C,D$ and $E$ as shown in Table (\ref{tabl}). Symbolically, these 15 textures can be represented as\\
\begin{equation}
\nonumber
A_{1}=
\begin{pmatrix}
0 & 0  & X \\
0 & X  & X \\
X & X  & X\\
\end{pmatrix}
, A_{2}=
\begin{pmatrix}
0 &  X & 0 \\
X & X  & X \\
0 & X  & X\\
\end{pmatrix}
, A_{3}=
\begin{pmatrix}
0 &  X & X \\
X & 0  & X \\
X & X  & X\\
\end{pmatrix},
\end{equation}
\begin{equation}
\nonumber
A_{4}=
\begin{pmatrix}
0 & X  & X \\
X & X  & 0 \\
X & 0  & X\\
\end{pmatrix}
,A_{5}=
\begin{pmatrix}
0 &  X & X \\
X & X  & X \\
X & X  & 0\\
\end{pmatrix};
\end{equation}

\begin{equation}
\nonumber
B_{1}=
\begin{pmatrix}
X & 0  & 0 \\
0 & X  & X \\
0 & X  & X\\
\end{pmatrix}
,B_{2}=
\begin{pmatrix}
X & 0  & X \\
0 & 0  & X \\
X & X  & X\\
\end{pmatrix}
,B_{3}=
\begin{pmatrix}
X &  0 & X \\
0 & X  & 0 \\
X & 0  & X\\
\end{pmatrix}
,B_{4}=
\begin{pmatrix}
X &  0 & X \\
0 & X  & X \\
X & X  & 0\\
\end{pmatrix};
\end{equation}

\begin{equation}
\nonumber
C_{1}=
\begin{pmatrix}
X & X  & 0 \\
X & 0  & X \\
0 & X  & X\\
\end{pmatrix}
,C_{2}=
\begin{pmatrix}
X & X  & 0 \\
X & X  & 0 \\
0 & 0  & X\\
\end{pmatrix}
,C_{3}=
\begin{pmatrix}
X &  X & 0 \\
X & X  & X \\
0 & X  & 0\\
\end{pmatrix};
\end{equation}

\begin{equation}
\nonumber
D_{1}=
\begin{pmatrix}
X & X  & X \\
X & 0  & 0 \\
X & 0  & X\\
\end{pmatrix}
,D_{2}=
\begin{pmatrix}
X & X  & X \\
X & 0  & X \\
X & X  & 0\\
\end{pmatrix};
\end{equation}

\begin{equation}
\nonumber
E_1=
\begin{pmatrix}
X & X  & X \\
X & X  & 0 \\
X & 0  & 0\\
\end{pmatrix},
\end{equation}
~~~\\

\noindent where “$X$” denotes the non-zero element of inverse neutrino mass matrix. In the present work, we investigate the phenomenological consequences of these 15 possible two-zero textures in $M_{\nu}^{-1}$ within the paradigm of LMA and LMA-D descriptions of neutrino oscillation phenomenon.\\
\noindent In general, two-zero texture of $M_{\nu}^{-1}$ result in constraining equations
\begin{eqnarray}
(M^{-1}_{\nu})_{pq}=0;\hspace{0.5cm}
(M^{-1}_{\nu})_{rs}=0
\label{5}
\end{eqnarray}
where $p,q,r$ and $s$ can take value 1,2,3 such that $p\leq q$, $r\leq s$.\\
Using Eqn. (\ref{3}), the constraints described by Eqn. (\ref{5}) can be written as
\begin{equation}
\lambda_{1}{U^{-1}}_{1p}{U^{-1}}_{1q}+\lambda_{2}{U^{-1}}_{2p}{U^{-1}}_{2q}+\lambda_{3}{U^{-1}}_{3p}{U^{-1}}_{3q}=0
\label{6},
\end{equation}
and
\begin{equation}
\lambda_{1}{U^{-1}}_{1r}{U^{-1}}_{1s}+\lambda_{2}{U^{-1}}_{2r}{U^{-1}}_{2s}+\lambda_{3}{U^{-1}}_{3r}{U^{-1}}_{3s}=0
\label{7},  
\end{equation}
where
\begin{eqnarray}
\nonumber
\lambda_{1}=\dfrac{1}{m_{1}} , \hspace{5mm}
\lambda_{2}=\dfrac{e^{-2i\alpha}}{m_{2}} , \hspace{5mm} 
\lambda_{3}=\dfrac{e^{-2i(\beta+\delta)}}{m_{3}}.
\end{eqnarray}
\begin{table}[t]
\begin{center}
\begin{tabular}{|c|c|c|c|c|} 
\hline
Class & Textures & $(M^{-1}_{\nu})_{pq}=0$ & $(M^{-1}_{\nu})_{rs}=0$\\
\hline
\multirow{5}{*}{A} & $A_1$ & $(M^{-1}_{\nu})_{11}$ & $(M^{-1}_{\nu})_{12}$ \\ 
& $A_2$ & $(M^{-1}_{\nu})_{11}$ &  $(M^{-1}_{\nu})_{13}$ \\ 
& $A_3$ & $(M^{-1}_{\nu})_{11}$ &  $(M^{-1}_{\nu})_{22}$ \\
& $A_4$ & $(M^{-1}_{\nu})_{11}$ &  $(M^{-1}_{\nu})_{23}$ \\
& $A_5$ & $(M^{-1}_{\nu})_{11}$ &  $(M^{-1}_{\nu})_{33}$ \\
\hline
\multirow{4}{*}{B} & $B_1$ & $(M^{-1}_{\nu})_{12}$ & $(M^{-1}_{\nu})_{13}$ \\ 
& $B_2$ & $(M^{-1}_{\nu})_{12}$ &   $(M^{-1}_{\nu})_{22}$\\ 
& $B_3$ & $(M^{-1}_{\nu})_{12}$ &  $(M^{-1}_{\nu})_{23}$ \\
& $B_4$ & $(M^{-1}_{\nu})_{12}$ &  $(M^{-1}_{\nu})_{33}$\\
\hline
\multirow{3}{*}{C} & $C_1$ & $(M^{-1}_{\nu})_{13}$ & $(M^{-1}_{\nu})_{22}$ \\ 
& $C_2$ & $(M^{-1}_{\nu})_{13}$ &  $(M^{-1}_{\nu})_{23}$\\ 
& $C_3$ & $(M^{-1}_{\nu})_{13}$ &  $(M^{-1}_{\nu})_{33}$\\
\hline
\multirow{2}{*}{D} & $D_1$ & $(M^{-1}_{\nu})_{22}$ & $(M^{-1}_{\nu})_{23}$\\ 
& $D_2$ & $(M^{-1}_{\nu})_{22}$ & $(M^{-1}_{\nu})_{33}$\\ 
\hline
\multirow{1}{*}{E} & $E_1$ & $(M^{-1}_{\nu})_{23}$ & $(M^{-1}_{\nu})_{33}$\\ 
\hline
\end{tabular}
         \caption{Fifteen possible two-zero texture patterns with corresponding first and second zero.}
        \label{tabl}
\end{center}
\end{table}
\noindent We solve Eqns. (\ref{6}) and (\ref{7}) for two mass ratios $\left(\dfrac{m_1}{m_2}e^{-2i\alpha},\hspace{0.05cm}\dfrac{m_1}{m_3}e^{-2i(\beta+\delta)}\right)$
\begin{eqnarray}
&&\dfrac{m_1}{m_2}e^{-2i\alpha}=\dfrac{{U^{-1}}_{3r}{U^{-1}}_{3s}{U^{-1}}_{1p}{U^{-1}}_{1q}-{U^{-1}}_{3p}{U^{-1}}_{3q}{U^{-1}}_{1r}{U^{-1}}_{1s}}{{U^{-1}}_{3p}{U^{-1}}_{3q}{U^{-1}}_{2r}{U^{-1}}_{2s}-{U^{-1}}_{2p}{U^{-1}}_{2q}{U^{-1}}_{3r}{U^{-1}}_{3s}},
\label{8}\\
&&\dfrac{m_1}{m_3}e^{-2i(\beta+\delta)}=\dfrac{{U^{-1}}_{1r}{U^{-1}}_{1s}{U^{-1}}_{2p}{U^{-1}}_{2q}-{U^{-1}}_{1p}{U^{-1}}_{1q}{U^{-1}}_{2r}{U^{-1}}_{2s}}{{U^{-1}}_{3p}{U^{-1}}_{3q}{U^{-1}}_{2r}{U^{-1}}_{2s}-{U^{-1}}_{2p}{U^{-1}}_{2q}{U^{-1}}_{3r}{U^{-1}}_{3s}}.
\label{9}
\end{eqnarray}
The absolute values of mass ratios are given by
\begin{equation}
\frac{m_1}{m_2}=\left|\dfrac{{U^{-1}}_{3r}{U^{-1}}_{3s}{U^{-1}}_{1p}{U^{-1}}_{1q}-{U^{-1}}_{3p}{U^{-1}}_{3q}{U^{-1}}_{1r}{U^{-1}}_{1s}}{{U^{-1}}_{3p}{U^{-1}}_{3q}{U^{-1}}_{2r}{U^{-1}}_{2s}-{U^{-1}}_{2p}{U^{-1}}_{2q}{U^{-1}}_{3r}{U^{-1}}_{3s}}\right|,
\label{10}
\end{equation}
\begin{equation}
\frac{m_1}{m_3}=\left|\dfrac{{U^{-1}}_{1r}{U^{-1}}_{1s}{U^{-1}}_{2p}{U^{-1}}_{2q}-{U^{-1}}_{1p}{U^{-1}}_{1q}{U^{-1}}_{2r}{U^{-1}}_{2s}}{{U^{-1}}_{3p}{U^{-1}}_{3q}{U^{-1}}_{2r}{U^{-1}}_{2s}-{U^{-1}}_{2p}{U^{-1}}_{2q}{U^{-1}}_{3r}{U^{-1}}_{3s}}\right|,
\label{11}
\end{equation}

\begin{table}[t]
    \centering
    \begin{tabular}{|l|l|l|}
    \hline
     Class   & Texture & ~~~~~~~~~~~~~~~~~~~~~~~~~~~~~~~~~~~~~Mass ratios $\left(R_{12},R_{13}\right)$  \\
     \hline
      & \multirow{2}{*}{$A_1$}  & $\dfrac{m_1}{m_2}e^{-2i\alpha}= \dfrac{c_{12}\left(c_{23}e^{i\delta}s_{12}s_{13}+c_{12}s_{23} \right)}{s_{12}\left(c_{12}c_{23}e^{i\delta}s_{13}-s_{12}s_{23}\right)}$ \\
    &&$\dfrac{m_1}{m_3}e^{-i(2\beta+\delta)}= \dfrac{c_{12}c^2_{13}c_{23}}{s_{13}\left(-c_{12}c_{23}e^{i\delta}s_{13}+s_{12}s_{23}\right)}$\\
    \cline{2-3}
   \multirow{5}{*}{A} & \multirow{2}{*}{$A_2$} & $\dfrac{m_1}{m_2}e^{-2i\alpha}=-\dfrac{c_{12}\left(c_{12}c_{23}-e^{i\delta}s_{12}s_{13}s_{23}\right)}{s_{12}\left(c_{23}s_{12}+c_{12}e^{i\delta}s_{13}s_{23}\right)}$\\
    && $\dfrac{m_1}{m_3}e^{-i(2\beta+\delta)}= \dfrac{c_{12}c^2_{13}s_{23}}{s_{13}\left(c_{23}s_{12}+c_{12}e^{i\delta}s_{13}s_{23}\right)}$\\ \cline{2-3}
    
     & \multirow{2}{*}{$A_3$} & $\dfrac{m_1}{m_2}e^{-2i\alpha}=-\dfrac{\left(c_{23}e^{i\delta}s_{12}s_{13}+c_{12}s_{23}\right)\left(c_{23}e^{i\delta}s_{12}s_{13}+c_{12}(s^2_{13}-c^2_{13})s_{23}\right)}{c^{2}_{12}c^{2}_{23}e^{2i\delta}s^{2}_{13}-2c_{12}c_{23}e^{i\delta}s_{12}s^{2}_{13}s_{23}+s^{2}_{12}(s^{4}_{13}-c^{4}_{13})s_{23}}$ \\
      & & $\dfrac{m_1}{m_3}e^{-i(2\beta+\delta)}= \dfrac{c^{2}_{13}c_{23}\left(c_{23}e^{i\delta}(c^{2}_{12}-s^{2}_{12})-2c_{12}s_{12}s_{13}s_{23}\right)}{-c^{2}_{12}c^{2}_{23}e^{2i\delta}s^{2}_{13}+2c_{12}c_{23}e^{i\delta}s_{12}s^{2}_{13}s_{23}+s^{2}_{12}(s^{4}_{13}-c^{4}_{13})s_{23}}$ \\ \cline{2-3}
       & \multirow{2}{*}{$A_4$} & $\dfrac{m_1}{m_2}e^{-2i\alpha}=\dfrac{c^2_{12}c^2_{23}(s^4_{13}-c^4_{13})-2c_{12}c_{23}e^{i\delta}s_{12}s^3_{13}s_{23}+e^{2i\delta}s^2_{12}s^2_{13}s^2_{23}}{c^2_{13}c^2_{23}s^2_{12}-s^2_{13}\left(c_{23}s_{12}s_{13}+c_{12}e^{i\delta}s_{23}\right)^2}$ \\
       & &$ \dfrac{m_1}{m_3}e^{-i(2\beta+\delta)}=-\dfrac{c^2_{13}s_{23}\left(2c_{12}c_{23}s_{12}s_{13}+e^{i\delta}(c^2_{12}-s^2_{12})s_{23}\right)}{-c^2_{13}c^2_{23}s^2_{12}+s^2_{13}\left(c_{23}s_{12}s_{13}+c_{12}e^{i\delta}s_{23}\right)^2}$ \\ \cline{2-3}
       & \multirow{2}{*}{$A_5$} & $\dfrac{m_1}{m_2}e^{-2i\alpha}=\dfrac{-c_{23}e^{2i\delta}s^2_{12}s^2_{13}s_{23}+c^2_{12}c_{23}(s^{4}_{13}-c^{4}_{13})s_{23}+c_{12}e^{i\delta}s_{12}s^3_{13}(c^{2}_{13}-s^{2}_{13})}{c^2_{12}c_{23}e^{2i\delta}s^2_{13}s_{23}+c_{23}s^2_{12}(c^{4}_{13}-s^{4}_{13})s_{23}+c_{12}e^{i\delta}s_{12}s^3_{13}(c^2_{23}-s^2_{23})}$\\
       &&$\dfrac{m_1}{m_3}e^{-i(2\beta+\delta)}=-\dfrac{-c^2_{13}\left(c_{23}e^{i\alpha}(c^2_{12}-s^2_{12})s_{23}+c_{12}s_{12}s_{13}(c^2_{23}-s^2_{23})\right)}{c^2_{12}c_{23}e^{2i\delta}s^2_{13}s_{23}+c_{23}s^2_{12}(c^{4}_{13}-s^{4}_{13})s_{23}+c_{12}e^{i\delta}s_{12}s^3_{13}(c^2_{23}-s^2_{23})}$\\
    \hline
    \end{tabular}
    \caption{Mass ratios for class $A$.}
    \label{tab2}
\end{table}
~~~~~~~~~~~~~~~~~~~~~~~~~~~~~~~~~~~\
\begin{table}[htb]
    \centering
    \begin{tabular}{|l|l|l|}
    \hline
    Class & Texture & ~~~~~~~~~~~~~~~~~~~~~~~~~~~~~~~~~~~~~Mass ratios $\left(R_{12},R_{13}\right)$\\
    \hline
    & $B_1$  & $ \dfrac{m_1}{m_2}e^{-2i\alpha}=1$ \\
    &&$\dfrac{m_1}{m_3}e^{-i(2\beta+\delta)}=1$\\ \cline{2-3}
  \multirow{4}{*}{B}  &\multirow{2}{*}{$B_2$}& $\dfrac{m_1}{m_2}e^{-2i\alpha}=-\dfrac{\left(c_{23}e^{i\delta}s_{12}+c_{12}s_{13}s_{23}\right)\left(c_{23}e^{i\delta}s_{12}s_{13}+c_{12}s_{23}\right)}{\left(c_{12}c_{23}e^{i\delta}-s_{12}s_{23}s_{13}\right)\left(c_{12}c_{23}e^{i\delta}s_{13}-s_{12}s_{23}\right)}$ \\
    && $\dfrac{m_1}{m_3}e^{-i(2\beta+\delta)}=\dfrac{c_{23}\left(c_{23}e^{i\delta}s_{12}+c_{12}s_{13}s_{23}\right)}{s_{23}\left(c_{12}c_{23}e^{i\delta}s_{13}-s_{12}s_{23}\right)}$ \\ \cline{2-3}
     &\multirow{2}{*}{$B_3$}& $\dfrac{m_1}{m_2}e^{-2i\alpha}=\dfrac{\left(c_{23}e^{i\delta}s_{12}+c_{12}s_{13}s_{23}\right)\left(c_{12}c_{23}-e^{i\delta}s_{12}s_{13}s_{23}\right)}{\left(c_{23}s_{12}+c_{12}e^{i\delta}s_{13}s_{23}\right)\left(c_{12}c_{23}e^{i\delta}-s_{12}s_{13}s_{23}\right)}$\\
    && $ \dfrac{m_1}{m_3}e^{-i(2\beta+\delta)}=\dfrac{c_{23}e^{i\delta}s_{12}+c_{12}s_{13}s_{23}}{c_{23}s_{12}+c_{12}e^{i\delta}s_{13}s_{23}}$ \\ \cline{2-3}
    &\multirow{2}{*}{$B_4$}& $ \dfrac{m_1}{m_2}e^{-2i\alpha}=-\dfrac{\left(c^2_{12}c^2_{23}s_{13}s_{23}+e^{2i\delta}s^{2}_{12}s_{13}s^{3}_{23}+c_{12}c_{23}e^{i\delta}s_{12}(c^{2}_{13}c^{2}_{23}-s^{2}_{13}s^{2}_{23})\right)}{c^{2}_{23}s^{2}_{12}s_{13}s_{23}+c^{2}_{12}e^{2i\delta}s_{13}s^{3}_{23}-c_{12}c_{23}e^{i\delta}s_{12}(c^{2}_{13}c^{2}_{23}-s^{2}_{13}s^{2}_{23})}$\\
    && $\dfrac{m_1}{m_3}e^{-i(2\beta+\delta)}=\dfrac{\left(c_{12}c_{23}e^{2i\delta}s_{12}s^{2}_{23}+e^{i\delta}(c^{2}_{12}-s^{2}_{12})s_{13}s^{3}_{23}+c_{12}c_{23}s_{12}s^{2}_{13}(c^{2}_{23}+2s^{2}_{23})\right)}{c^{2}_{23}s^{2}_{12}s_{13}s_{23}+c^{2}_{12}e^{2i\delta}s_{13}s^{3}_{23}-c_{12}c_{23}e^{i\delta}s_{12}(c^{2}_{13}c^{2}_{23}-s^{2}_{13}s^{2}_{23})}$\\
    \hline
    
    \end{tabular}
    \caption{Mass ratios for class $B$.}
    \label{tab3}
\end{table}

\begin{table}[h!]
    \centering
    \begin{tabular}{|l|l|l|}
    \hline
      Class   & Texture & ~~~~~~~~~~~~~~~~~~~~~~~~~~~~~~~~~~~~~~~~~~~~~~~~Mass ratios $\left(R_{12},R_{13}\right)$\\
      \hline
      & \multirow{2}{*}{$C_1$} & $\dfrac{m_1}{m_2}e^{-2i\alpha}=-\dfrac{\left(c^{3}_{23}e^{2i\delta}s^{2}_{12}s_{13}+c^{2}_{12}c_{23}s_{13}s^{2}_{23}+c_{12}e^{i\delta}s_{12}s_{23}(2c^{2}_{23}s^{2}_{13}-c^{2}_{13}s^{2}_{23})\right)}{c^{2}_{12}c^{3}_{23}s_{13}e^{2i\delta}+c_{23}s^{2}_{12}s_{13}s^{2}_{23}+c_{12}e^{i\delta}s_{12}s_{23}(-2c^{2}_{23}s^{2}_{13}+c^{2}_{13}s^{2}_{23})}$ \\ 
     && $\dfrac{m_1}{m_3}e^{-i(2\beta+\delta)}=-\dfrac{\left(c^{3}_{23}e^{i\delta}(c^{2}_{12}-s^{2}_{12})s_{13}+c_{12}c^{2}_{23}e^{2i\delta}s_{12}s_{23}+c_{12}s_{12}s^{2}_{13}s_{23}(2c^{2}_{23}+s^{2}_{23})\right)}{c^{2}_{12}c^{3}_{23}s_{13}e^{2i\delta}+c_{23}s^{2}_{12}s_{13}s^{2}_{23}+c_{12}e^{i\delta}s_{12}s_{23}(-2c^{2}_{23}s^{2}_{13}+c^{2}_{13}s^{2}_{23})}$\\ \cline{2-3}
     \multirow{3}{*}{C} &\multirow{2}{*}{$C_2$}& $ \dfrac{m_1}{m_2}e^{-2i\alpha}=\dfrac{\left(c_{12}c_{23}s_{13}-e^{i\delta}s_{12}s_{23}\right)\left(c_{23}e^{i\delta}s_{12}s_{13}+c_{12}s_{23}\right)}{\left(c_{23}s_{12}s_{13}+c_{12}e^{i\delta}s_{23}\right)\left(c_{12}c_{23}e^{i\delta}s_{13}-s_{12}s_{23}\right)}$\\
     && $\dfrac{m_1}{m_3}e^{-i(2\beta+\delta)}=-\dfrac{\left(-c_{12}c_{23}s_{13}+e^{i\delta}s_{12}s_{23}\right)}{c_{12}c_{23}e^{i\delta}s_{13}-s_{12}s_{23}}$ \\ \cline{2-3}
     &\multirow{2}{*}{$C_3$}& $\dfrac{m_1}{m_2}e^{-2i\alpha}=\dfrac{\left(c_{12}c_{23}s_{13}-e^{i\delta}s_{12}s_{23}\right)\left(c_{12}c_{23}-e^{i\delta}s_{12}s_{13}s_{23}\right)}{\left(c_{23}s_{12}s_{13}+c_{12}e^{i\delta}s_{23}\right)\left(c_{23}s_{12}+c_{12}e^{i\delta}s_{13}s_{23}\right)}$ \\
     &&$\dfrac{m_1}{m_3}e^{-i(2\beta+\delta)}=\dfrac{s_{23}\left(c_{12}c_{23}s_{13}-e^{i\delta}s_{12}s_{23}\right)}{c_{23}\left(c_{23}s_{12}+c_{12}e^{i\delta}s_{13}s_{23}\right)}$\\ \hline
     &\multirow{2}{*}{$D_1$}& $\dfrac{m_1}{m_2}e^{-2i\alpha}=-\dfrac{s_{12}\left(c_{23}e^{i\delta}s_{12}+c_{12}s_{13}s_{23}\right)}{c_{12}\left(c_{12}c_{23}e^{i\delta}-s_{12}s_{23}s_{13}\right)}$\\
      \multirow{2}{*}{D}&& $\dfrac{m_1}{m_3}e^{-i(2\beta+\delta)}=\dfrac{s_{13}\left(c_{23}s_{12}+e^{-i\delta}c_{12}s_{13}s_{23}\right)}{c_{12}c^{2}_{13}s_{23}}$ \\ \cline{2-3}
   &\multirow{2}{*}{$D_2$}& $\dfrac{m_1}{m_2}e^{-2i\alpha}=-\dfrac{s_{12}\left(2c_{12}c_{23}s_{13}s_{23}+e^{i\delta}s_{12}(c^{2}_{23}-s^{2}_{23})\right)}{c_{12}\left(-2c_{23}s_{12}s_{13}s_{23}+c_{12}e^{i\delta}(c^{2}_{23}-s^{2}_{23})\right)}$ \\
     && $ \dfrac{m_1}{m_3}e^{-i(2\beta+\delta)}=-\dfrac{s_{13}\left(c^{2}_{23}s_{13}(c^{2}_{12}-s^{2}_{12})-2c_{12}c_{23}s_{12}s_{23}(e^{-i\delta}s^{2}_{13}+e^{i\delta})+s_{13}s^{2}_{23}(s^{2}_{12}-c^{2}_{12})\right)}{c_{12}c^{2}_{13}\left(c_{12}c^{2}_{23}e^{i\delta}-2c_{23}s_{12}s_{13}s_{23}-c_{12}e^{i\delta}s^{2}_{23}\right)}$ \\ \hline
   \multirow{1}{*}{$E$}  & $E_1$ & $\dfrac{m_1}{m_2}e^{-2i\alpha}=\dfrac{s_{12}\left(c_{12}c_{23}s_{13}-e^{i\delta}s_{12}s_{23}\right)}{c_{12}\left(c_{23}s_{12}s_{13}+c_{12}e^{i\delta}s_{23}\right)}$\\
     &&$ \dfrac{m_1}{m_3}e^{-i(2\beta+\delta)}=\dfrac{s_{13}\left(-e^{-i\delta}c_{12}c_{23}s_{13}+s_{12}s_{23}\right)}{c_{12}c_{23}c^{2}_{13}}$\\
     \hline
    \end{tabular}
    \caption{Mass ratios for class $C,D$ and $E$.}
    \label{tab4}
\end{table}
\noindent and two Majorana phases are obtained as
\begin{eqnarray}
&&\alpha=-\dfrac{1}{2} Arg\left(\dfrac{{U^{-1}}_{3r}{U^{-1}}_{3s}{U^{-1}}_{1p}{U^{-1}}_{1q}-{U^{-1}}_{3p}{U^{-1}}_{3q}{U^{-1}}_{1r}{U^{-1}}_{1s}}{{U^{-1}}_{3p}{U^{-1}}_{3q}{U^{-1}}_{2r}{U^{-1}}_{2s}-{U^{-1}}_{2p}{U^{-1}}_{2q}{U^{-1}}_{3r}{U^{-1}}_{3s}}\right),
\label{12}\\
&&\beta=-\dfrac{1}{2} Arg\left(\dfrac{{U^{-1}}_{1r}{U^{-1}}_{1s}{U^{-1}}_{2p}{U^{-1}}_{2q}-{U^{-1}}_{1p}{U^{-1}}_{1q}{U^{-1}}_{2r}{U^{-1}}_{2s}}{{U^{-1}}_{3p}{U^{-1}}_{3q}{U^{-1}}_{2r}{U^{-1}}_{2s}-{U^{-1}}_{2p}{U^{-1}}_{2q}{U^{-1}}_{3r}{U^{-1}}_{3s}}\right)-\delta.
\label{13}
\end{eqnarray}\\
It is to be noted that the mass ratios $\left(\dfrac{m_1}{m_2}e^{-2i\alpha},\hspace{0.05cm}\dfrac{m_1}{m_3}e^{-2i(\beta+\delta)}\right)$ are different for each texture as they depend on the position of zero in $M^{-1}_{\nu}$. For example, in case of $A_{1}$ texture
\begin{eqnarray}\nonumber
p=1, \hspace{0.5cm} q=1, \hspace{0.5cm} r=1, \hspace{0.5cm} s=2,
\end{eqnarray}
therefore, Eqns. (\ref{8}) and (\ref{9}) become
\begin{eqnarray}\nonumber
&&\dfrac{m_1}{m_2}e^{-2i\alpha}=\dfrac{{U^{-1}}_{31}{U^{-1}}_{32}{U^{-1}}_{11}{U^{-1}}_{11}-{U^{-1}}_{31}{U^{-1}}_{31}{U^{-1}}_{11}{U^{-1}}_{12}}{{U^{-1}}_{31}{U^{-1}}_{31}{U^{-1}}_{21}{U^{-1}}_{22}-{U^{-1}}_{21}{U^{-1}}_{21}{U^{-1}}_{31}{U^{-1}}_{32}},
\\ \nonumber
&&\dfrac{m_1}{m_3}e^{-2i(\beta+\delta)}=\dfrac{{U^{-1}}_{11}{U^{-1}}_{12}{U^{-1}}_{21}{U^{-1}}_{21}-{U^{-1}}_{11}{U^{-1}}_{11}{U^{-1}}_{21}{U^{-1}}_{22}}{{U^{-1}}_{31}{U^{-1}}_{31}{U^{-1}}_{21}{U^{-1}}_{22}-{U^{-1}}_{21}{U^{-1}}_{21}{U^{-1}}_{31}{U^{-1}}_{32}},
\end{eqnarray}
while for $B_{2}$ texture
\begin{eqnarray}\nonumber
p=1, \hspace{0.5cm} q=2, \hspace{0.5cm} r=2, \hspace{0.5cm} s=2,
\end{eqnarray}
resulting in mass ratios
\begin{eqnarray}\nonumber
&&\dfrac{m_1}{m_2}e^{-2i\alpha}=\dfrac{{U^{-1}}_{32}{U^{-1}}_{32}{U^{-1}}_{11}{U^{-1}}_{12}-{U^{-1}}_{31}{U^{-1}}_{32}{U^{-1}}_{12}{U^{-1}}_{12}}{{U^{-1}}_{31}{U^{-1}}_{32}{U^{-1}}_{22}{U^{-1}}_{22}-{U^{-1}}_{21}{U^{-1}}_{22}{U^{-1}}_{32}{U^{-1}}_{32}},
\label{81}\\ \nonumber
&&\dfrac{m_1}{m_3}e^{-2i(\beta+\delta)}=\dfrac{{U^{-1}}_{12}{U^{-1}}_{12}{U^{-1}}_{21}{U^{-1}}_{22}-{U^{-1}}_{11}{U^{-1}}_{12}{U^{-1}}_{22}{U^{-1}}_{22}}{{U^{-1}}_{31}{U^{-1}}_{32}{U^{-1}}_{22}{U^{-1}}_{22}-{U^{-1}}_{21}{U^{-1}}_{22}{U^{-1}}_{32}{U^{-1}}_{32}}.
\label{91}
\end{eqnarray}
\noindent We have analytically solved the mass ratios for all possible two-zero textures which are shown in the Tables (\ref{tab2}-\ref{tab4}).\\
\noindent The two mass-squared differences $\Delta m_{21}^{2}=m_{2}^{2}-m_{1}^{2}$ and $\left|\Delta m_{32}^{2}\right|=m_{3}^{2}-m_{2}^{2}$ alongwith mass ratios $\dfrac{m_1}{m_2}e^{-2i\alpha}\equiv R_{12}$, $\dfrac{m_1}{m_3}e^{-i(2\beta+\delta)}\equiv R_{13}$ yield two values of neutrino mass $m_{1}$ given by
\begin{equation}
m_{1}^{a}=\left|R_{12}\right|\sqrt{\dfrac{\Delta m^2_{21}}{1-\left|R_{12}\right|^2}},\hspace{3mm} m_{1}^{b}=\left|R_{13}\right|\sqrt{\dfrac{\Delta m^2_{21}+|\Delta m^2_{32}|}{1-\left|R_{13}\right|^2}},
\label{14}
\end{equation}

\noindent respectively. In order to ensure the consistency of the formalism the two values ($m_{1}^{a},  m_{1}^{b}$) must be equal which results in mass ratio parameter
\begin{eqnarray}
\dfrac{\Delta m^{2}_{21}}{|\Delta m^{2}_{32}|}= \dfrac{|R_{13}|^{2}\left(1-|R_{12}|^{2}\right)}{|R_{12}|^{2}-|R_{13}|^{2}}\equiv R_{\nu}.
\label{15}
\label{R}
\end{eqnarray}
The $3\sigma$ experimental range of parameter $R_{\nu}$ defined in Eqn. (\ref{R}) is 0.02590 $< R_{\nu} <$ 0.03656. The allowed phenomenology of the model is obtained by restricting $R_{\nu}$ in the $3\sigma$ experimental range. The neutrino mass eigenvalues $m_{2}$ and $m_{3}$ can be obtained using mass square differences $\left(\Delta m_{21}^{2},\hspace{0.05cm}\Delta m_{32}^{2}\right)$ as
\begin{eqnarray}
&&m_{2}=\sqrt{m_{1}^{2}+\Delta m_{21}^{2}};\hspace{0.1cm}
m_{3}=\sqrt{m_{2}^{2}+\Delta m_{32}^{2}}\hspace{0.2cm}\text{for normal hierarchy (NH)   $(m_{1}<m_{2}<m_{3})$},\\ \nonumber
\text{and}\\ 
&&m_{2}=\sqrt{m_{1}^{2}+\Delta m_{21}^{2}};\hspace{0.1cm}
m_{1}=\sqrt{m_{3}^{2}+\Delta m_{32}^{2}-\Delta m_{21}^{2}}\hspace{0.2cm}\text{for inverted hierarchy (IH)   $(m_{3}<m_{1}<m_{2})$}.
\end{eqnarray}
\noindent Also, the effective Majorana mass which governs the neutrinoless double beta($0\nu\beta\beta$) decay process is given by 
\begin{equation}
|M_{ee}|=\left|\sum_{i}V^{2}_{ei}m_{i}\right|=\left|m_1c^{2}_{12}c^{2}_{13}+m_2s^{2}_{12}c^{2}_{13}e^{2i\alpha}+m_3s^{2}_{13}e^{2i\beta}\right|.
\label{21}
\end{equation}
\noindent The Jarlskog $CP$ invariant is defined as \cite{jarlskog1,jarlskog2}
\begin{equation*} \label{jcp}
J_{C P}=s_{23} c_{23} s_{12} c_{12} s_{13} c_{13}^{2} \sin \delta.
\end{equation*}

\begin{table}[h]
        \centering
        \begin{tabular}{|c|c|c|}
        \hline
    Parameters  & best fit$\pm 1\sigma$ range(NH)&best fit$\pm 1\sigma$ range(IH) \\
    \hline
     $\Delta m^2_{21}[10^{-5}$ eV$^2]$ & $7.50^{+0.22}_{-0.20}$& $7.50^{+0.22}_{-0.20}$\\
     \hline
     $\left|\Delta m^2_{31}\right|[10^{-3}$ eV$^2]$& $2.55^{+0.02}_{-0.03}$& $2.45^{+0.02}_{-0.03}$\\
    \hline
    $\theta^\circ_{12}$ &  $34.3\pm 1.0$& $34.3\pm 1.0$\\
    \hline
     $\theta_{23}^\circ$ & $49.26\pm0.79$& $49.46^{+0.60}_{-0.97}$\\
     \hline
     $\theta_{13}^\circ$ & $8.53^{+0.13}_{-0.14}$&  $8.53^{+0.12}_{-0.14}$\\
    \hline
        \end{tabular}
        \caption{Global fit values of neutrino oscillation parameters\cite{deSalas:2020pgw}.}
        \label{tab5}
    \end{table}
\noindent In the numerical analysis, to study LMA phenomenology of the model, we have randomly generated the known neutrino oscillation parameters such as $\theta_{ij}$ $(i,j=1,2,3$; $i<j)$ and $\Delta m_{ij}^{2}$ $(i>j)$ using Gaussian distribution within allowed experimental range shown in Table (\ref{tab5}). However, in order to study the viability of the model under LMA-D solution, $\theta_{12}$ is randomly generated  using the uniform distribution within the range (53.71$^\circ$-58.37$^\circ$)\cite{Miranda:2007zza,Vishnudath:2019eiu,Escrihuela:2009up,Farzan:2017xzy}.
\section{LMA and LMA-D phenomenology}
In this section, we have investigated the phenomenology of all possible two-zero textures of $M_{\nu}^{-1}$ under the paradigm of LMA and LMA-D solutions to the neutrino oscillation phenomenon.
\subsection{Class A}
\begin{figure}
    \centering
    \includegraphics[width=7.5cm,height=6.5cm]{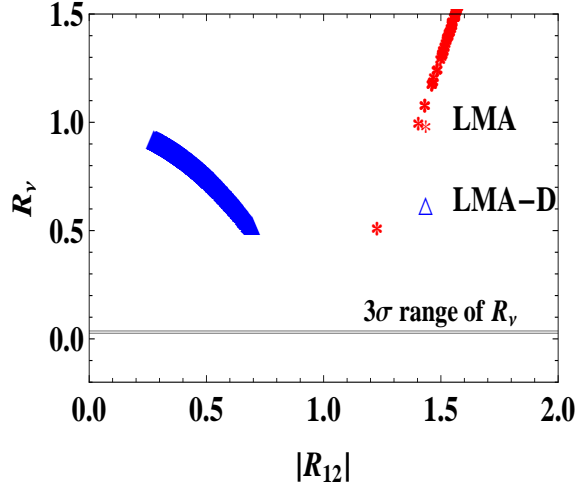}
    \caption{Correlation between $\left|R_{12}\right|$ and $R_{\nu}$ for texture $A_{1}$.}
    \label{figRnu}
\end{figure}

Class $A$ is disallowed for both LMA and LMA-D solutions. As a representative case, we have discussed the viability of $A_1$ texture in the following.
\noindent The mass ratios ($|R_{12}|, |R_{13}|$) for $A_1$ texture up-to first order in $s_{13}$ can be written as
\begin{eqnarray}
&&|R_{12}|\equiv\dfrac{m_{1}}{m_{2}}\approx \dfrac{c_{12}^{2}}{s_{12}^{2}}+\dfrac{c_{12}c_{23}\cos{\delta}}{s_{12}^{3}s_{23}}s_{13},
\label{19}\\
&&|R_{13}|\equiv\dfrac{m_{1}}{m_{3}}\approx \dfrac{c_{12}^{2}c_{23}^{2}\cos{\delta}}{s_{12}^{2}s_{23}^{2}}+\dfrac{c_{12}c_{23}\left(c_{12}^{2}c_{23}^{2}-4s_{12}^{2}s_{23}^{2}+3c_{12}^{2}c_{23}^{2}\cos{2\delta} \right)}{4 s_{12}^{3}s_{23}^{3}}.
\label{20}
\end{eqnarray}
\underline{\textbf{LMA Scenario:}} For $\delta$ in the range $0^{\circ}\leq \delta \leq 90^\circ$ or $270^{\circ}\leq \delta \leq 360^\circ$, Eqn. (\ref{19}) results in $\dfrac{m_{1}}{m_{2}}>1$. Furthermore, if $\delta$ lie in the range $90^{\circ}< \delta < 270^\circ$ and $\sin{\theta_{23}} \approx \cos{\theta_{23}}$, solar mass hierarchy requires $\sin{2{\theta_{12}}}<4\sin{\theta_{13}}$, however, from neutrino oscillation data (Table (\ref{tab5})) $\sin{2{\theta_{12}}}>4\sin{\theta_{13}}$ implying that $A_1$ texture is disallowed. 
\underline{\textbf{LMA-D Scenario:}} The parameter $(R_{\nu})$ up-to first order in $s_{13}$ can be written as
\begin{eqnarray}
R_{\nu}\approx\left(-1+\dfrac{c_{12}^{4}}{s_{12}^{4}}\right)+\dfrac{2c_{12}^{3}c_{23}\cos{\delta}}{s_{12}^{5}s_{23}}s_{13}.
\label{21}
\end{eqnarray}
 Using $\theta_{12}=55^\circ$ and $\delta$= $80^\circ$ $(190^\circ)$, numerical value of $R_{\nu}$ is found to be $0.74$ $(0.89)$ which lies outside the $3\sigma$ range of $R_{\nu}$. The above observations are, also, evident from Fig. (\ref{figRnu}).

\noindent Similar analysis can be done for all remaining textures in class $A$. Therefore, neutrino mass model, with two-zero textures in $M_{\nu}^{-1}$, wherein one of the texture zero is at $(1,1)$ place in $M_{\nu}^{-1}$ is disallowed.  

\subsection{Class B}
In class $B$, mass ratios for textures $B_1$ are given by
\begin{equation}
\nonumber
\left|R_{12}\right|\equiv\dfrac{m_{1}}{m_{2}}=1,
\hspace{3mm} \left|R_{13}\right|\equiv\dfrac{m_{1}}{m_{3}}=1,
\end{equation}
which result in degenerate neutrino masses (i.e. $m_1=m_2=m_3$) and are inconsistent with neutrino oscillation data. Also, for texture $B_3$, mass ratios are given by
\begin{eqnarray}
    &&|R_{12}|=\left|\dfrac{\left(c_{23}e^{i\delta}s_{12}+c_{12}s_{13}s_{23}\right)\left(c_{12}c_{23}-e^{i\delta}s_{12}s_{13}s_{23}\right)}{\left(c_{23}s_{12}+c_{12}e^{i\delta}s_{13}s_{23}\right)\left(c_{12}c_{23}e^{i\delta}-s_{12}s_{13}s_{23}\right)}\right|=1,\\
    && |R_{13}|=\left|\dfrac{c_{23}e^{i\delta}s_{12}+c_{12}s_{13}s_{23}}{c_{23}s_{12}+c_{12}e^{i\delta}s_{13}s_{23}}\right|=1, 
\end{eqnarray}
resulting in degenerate neutrino masses. Therefore, textures $B_1$ and $B_3$ are disallowed. In the following we have investigated the phenomenological consequences of $B_2$ and $B_4$ textures under LMA and LMA-D solutions.\\
The mass ratios ($|R_{12}|, |R_{13}|$) for texture $B_2$, up-to first order in $s_{13}$, can be written as
\begin{eqnarray}
\left|R_{12}\right|\equiv\dfrac{m_{1}}{m_{2}}\approx 1+\dfrac{2\cos{\delta}}{c_{12}c_{23}s_{12}s_{23}}s_{13},
\label{24}
\end{eqnarray}
\begin{eqnarray}
\left|R_{13}\right|\equiv\dfrac{m_{1}}{m_{3}}\approx \dfrac{c_{23}^{2}}{s_{23}^{2}}+\dfrac{c_{12}c_{23}\cos{\delta}}{s_{12}s_{23}^{3}}s_{13}.
\label{25}
\end{eqnarray}

\begin{figure}[h]
    \centering
    \includegraphics[width=16cm]{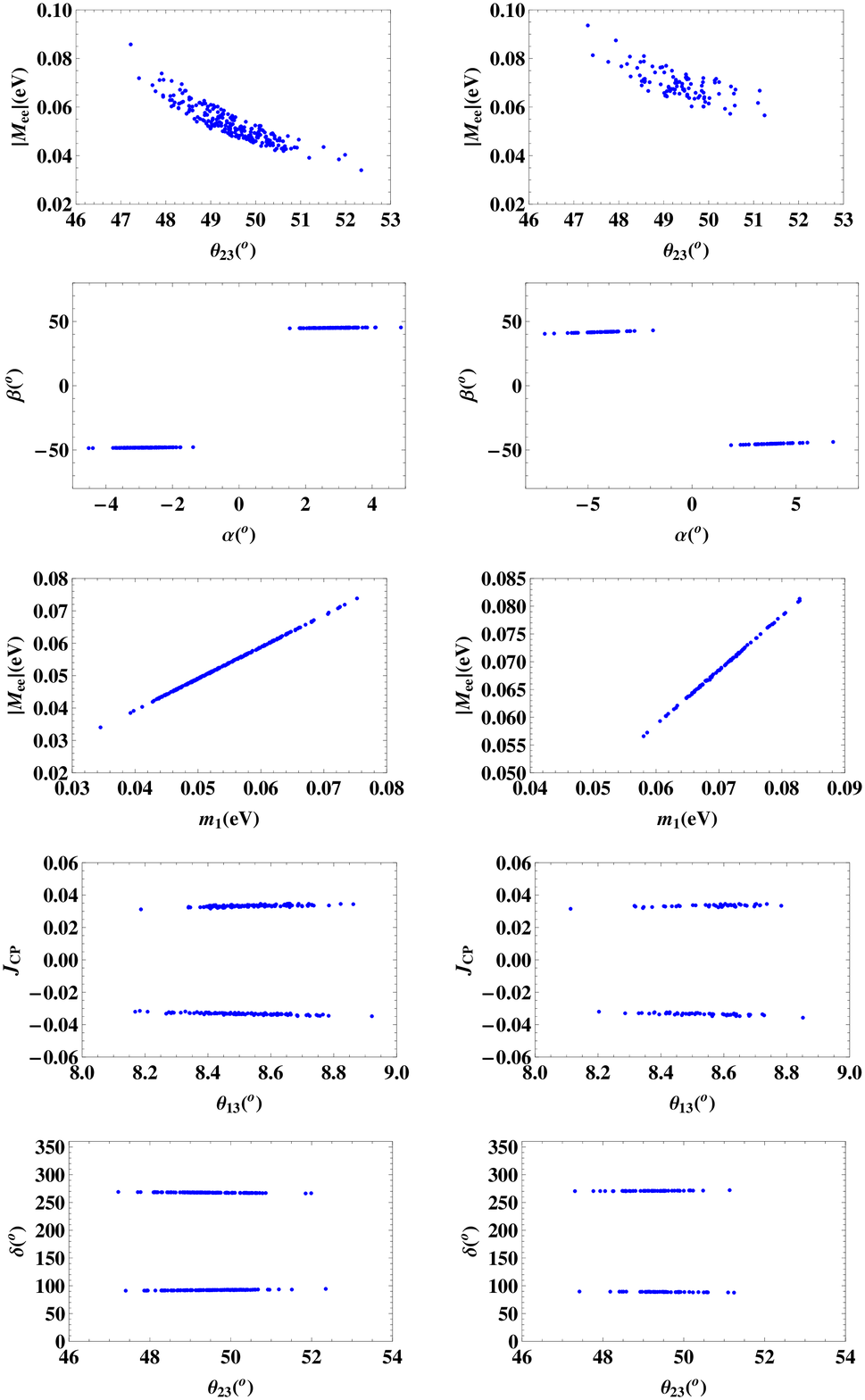}
    \caption{Correlation plots for textures $B_2$-NH(left panel) and $B_4$-IH(right panel) under LMA scenario.}
    \label{fig1}
\end{figure}

\noindent From Eqn. (\ref{24}), in order to satisfy the solar mass hierarchy $i.e$ $|R_{12}|\equiv\frac{m_1}{m_2}<1$,  $\delta$ should be in the range $90^\circ<\delta<270^{\circ}$. Also, texture $B_2$ predicts the normal hierarchical neutrino masses for $\theta_{23}$ above maximality ($\theta_{23}>45^o$) and $\cos\delta$ negative such that $\left|R_{13}\right|\equiv\dfrac{m_{1}}{m_{3}}<1$. The above observations are, also, depicted in the Fig. (\ref{fig1}). Similar analysis can also be done for texture $B_4$.\\
\noindent The texture $B_2$ ($B_4$) is allowed for both LMA and LMA-D solutions with normal (inverted) hierarchy. The allowed parameter space  for these textures are shown in Fig. (\ref{fig1}) as correlation plots amongst different parameters. Both these textures are found to have identical phenomenology under LMA and LMA-D solutions. In  Fig. (\ref{fig1}) we have shown the LMA scenario for $B_2$ (NH) and $B_4$ (IH) textures. In the left (right) panel we have depicted the correlation plots for $B_2$ ($B_4$) texture.  The atmospheric mixing angle $\theta_{23}$ is found to be above maximality for both the textures. The $CP$ violating phases $\alpha,\beta$ and $\delta$ are found to be sharply constrained. The Dirac-type $CP$ violating phase $\delta$ is found to be maximal (around $90^o$ and $270^o$) and the Jarlskog rephasing invariant $J_{CP}\ne 0$, thus, these textures are necessarily $CP$ violating. We can, also, appreciate the $CP$ violating nature of these textures analytically. For example, for texture $B_2$ with $\delta=0^o$ ($CP$ conserving scenario), we obtain $R_\nu$ to the first order in $s_{13}$ using Eqn. (\ref{R}) and values of mass ratios given in Table (\ref{tab3})
\begin{eqnarray}
\nonumber
R_{\nu}\approx\frac{2 c_{23} }{c_{12} s_{12} s_{23} \left(c_{23}^2-s_{23}^2\right)}s_{13}-\frac{2 c_{23}  s_{23} }{c_{12} s_{12} \left(c_{23}^2-s_{23}^2\right)}s_{13}.
\end{eqnarray}
Using the best-fit values given in Table (\ref{tab5}), $|R_\nu|\approx 1.7$ which is outside 3$\sigma$ range, thus, $\delta=0^o$ is disallowed implying $B_2$ texture is necessarily $CP$ violating. Similarly, for texture $B_4$, taking $\delta=0^{\circ}$ $R_{\nu}$ can, approximately, be written as
\begin{eqnarray}
\nonumber
R_{\nu}\approx\frac{s_{23}^5\left(2c_{23}^{4}c_{12}^2 -2c_{12}^2 s_{23}^2 c_{23}^2+2c_{23}^{2}s_{12}^{2}-s_{23}^{2}\right)}{c_{12}c_{23}^{5}s_{12}\left(s_{23}^{4}-c_{23}^{4}\right)}s_{13},
\end{eqnarray}
which again result in $|R_\nu|>1$ for best-fit values of the mixing angles. Similar analysis can be done for $CP$-violating textures of class C and D discussed in the following sections.

\noindent We have, also, obtained the implication of the model for neutrinoless double beta ($0\nu\beta\beta$) decay amplitude $|M_{ee}|$. It is evident from Fig. (\ref{fig1}) that there exist a lower bound on $|M_{ee}|$ in both the textures. For texture $B_2$ ($B_4$), $|M_{ee}|<0.03$ eV ($0.06$ eV). The prediction for $|M_{ee}|$ has, also, been tabulated in Table (\ref{tab7}).

\subsection{Class C}
\begin{figure}
    \centering
    \includegraphics[width=16cm]{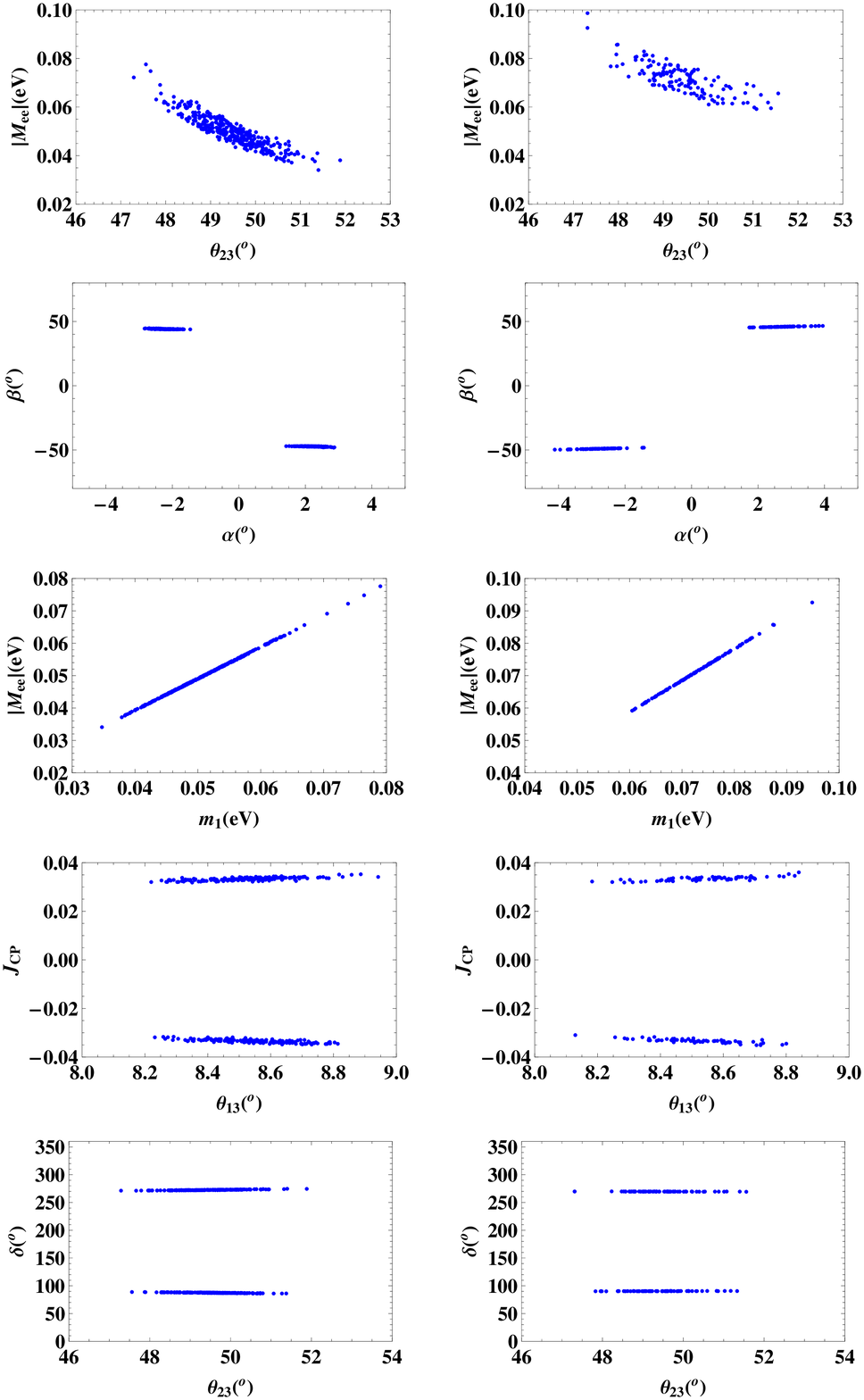}
    \caption{Correlation plots for textures $C_1$-NH(left panel) and $C_3$-IH(right panel) under LMA scenario.}
    \label{fig2}
\end{figure}

For texture $C_2$, the mass ratios $|R_{12}|$ and $|R_{13}|$ are equal to 1 resulting in degenerate neutrino masses which is in contradiction with neutrino oscillation data. Therefore, texture $C_2$ is disallowed. For textures $C_1$, the mass ratios $|R_{12}|$ and $|R_{13}|$, up-to first order in $s_{13}$, are given by
\begin{eqnarray}
 &&|R_{12}|\equiv\dfrac{m_{1}}{m_{2}}\approx1-\frac{c_{23}s_{13}\cos{\delta}}{c_{12}s_{12}s_{23}^3},
 \label{28}\\
&&|R_{13}|\equiv\dfrac{m_{1}}{m_{3}}\approx\frac{c_{23}^2}{s_{23}^2}-\frac{c_{12}c_{23}^3s_{13}\cos\delta}{s_{12}s_{23}^5}.
  \label{29}
  \end{eqnarray}
It can be seen from Eqn. (\ref{28}) that $\delta$ should lie in first and fourth quadrant to have $|R_{12}|\equiv\frac{m_1}{m_2}<1$. Also, for $\theta_{23}$ above maximality ($\theta_{23}>45^o$) and $\cos\delta$ positive the model predicts normal hierarchical neutrino masses. The above observations are, also, supplemented by Fig. (\ref{fig2}). Similar analysis can, also, be done for texture $C_3$. In fact, it is evident from Fig. (\ref{fig2}) that $C_3$ admit inverted hierarchical neutrino masses. Also, it is to be noted that the phenomenology of both the textures are found to be similar under LMA and LMA-D solutions. In Fig. (\ref{fig2}), we have shown the allowed parameter space considering LMA solution. The Majorana phases are sharply correlated and constrained to very narrow ranges giving a lower bound on $0\nu\beta\beta$ decay amplitude $|M_{ee}|$. For texture $C_1$ ($C_3$) $|M_{ee}|>0.02$ $(0.06)$ eV (Table (\ref{tab7})). The Dirac-type $CP$ violating phase $\delta$ is sharply constrained around $90^o$ and $270^o$, thus, allowing a maximal $CP$ violation. The Jarlskog rephasing invariant $J_{CP}$ is non-zero which is, also, shown in Fig. (\ref{fig2}).

\subsection{Class D}
For textures $D_1$, the mass ratios, up-to first order in $s_{13}$, are given by
\begin{eqnarray}
&&|R_{12}|\equiv\dfrac{m_{1}}{m_{2}}\approx \frac{s_{12}^2}{c_{12}^2}+\frac{s_{12}s_{23}s_{13}\cos\delta}{c_{12}^3c_{23}},
\label{30}\\
&&|R_{13}|\equiv\dfrac{m_{1}}{m_{3}}\approx \frac{\tan\theta_{12}}{\tan\theta_{23}}s_{13}.
\label{31}
\end{eqnarray}

\begin{figure}[h]
    \centering
    \includegraphics[width=16cm]{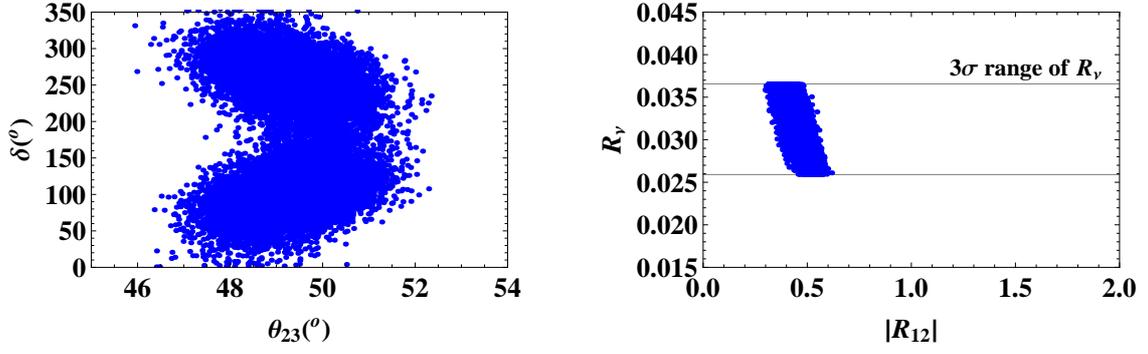}
    \caption{Correlation plots between $(\theta_{23}-\delta)$(left) and  $(\left|R_{12}\right|-R_{\nu})$(right) for $D_1$ texture under LMA solution with NH.}
    \label{fig3}
\end{figure}

\begin{figure}
    \centering
    \includegraphics[width=16cm]{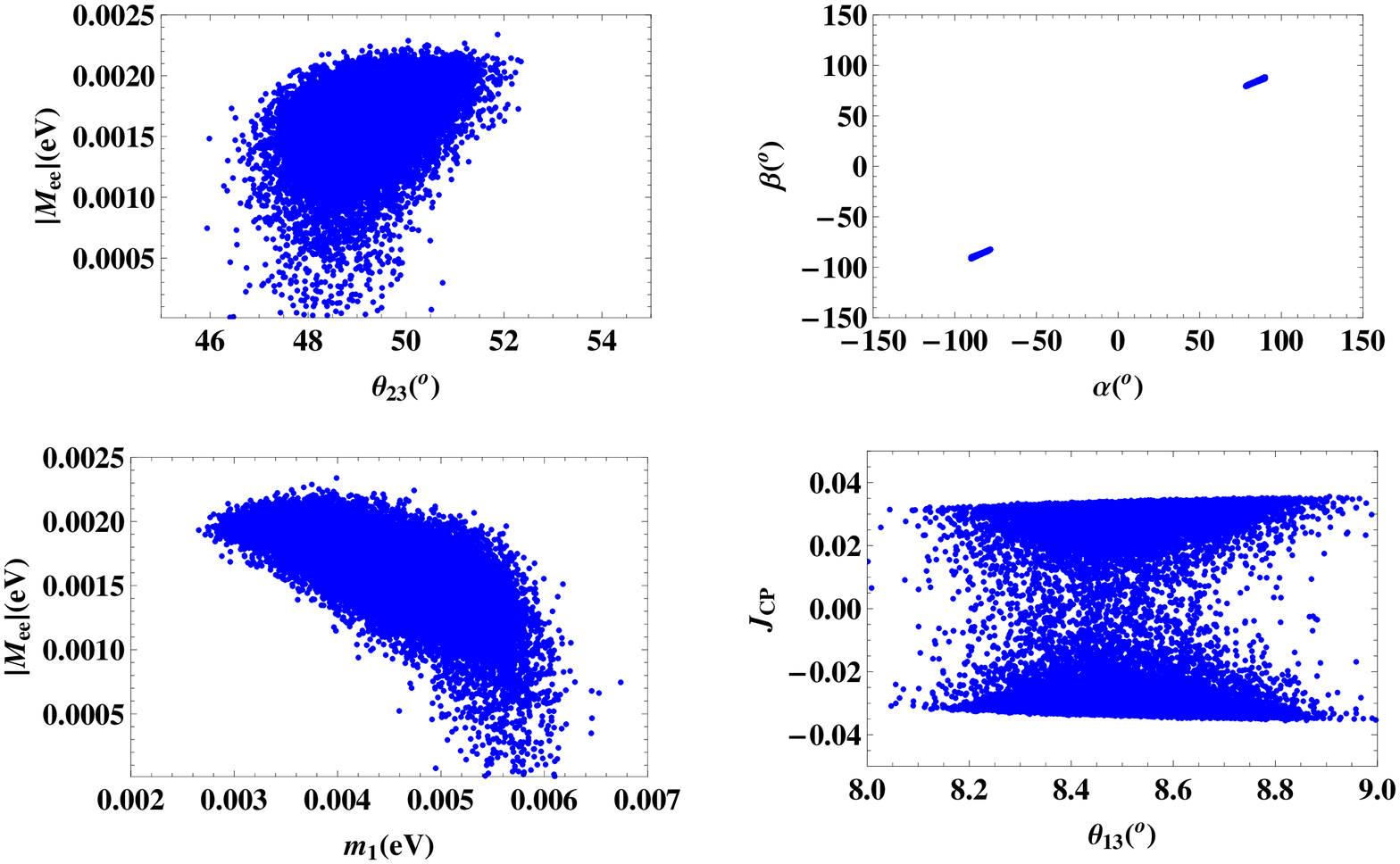}
    \caption{Correlation plots for $D_1$ texture under LMA solution with NH.}
    \label{fig4}
\end{figure}

\noindent For texture $D_1$, LMA-D solution is disallowed as $|R_{12}|>1$ (Eqn. (\ref{30})). However, for LMA solution, it is evident that $|R_{13}|<1$(Eqn. (\ref{31})) implying normal hierarchical neutrino masses. The above analytical observations are, also, supplemented by the correlation plots in Figs. (\ref{fig3}) and (\ref{fig4}). In addition, the Majorana phases are sharply constrained and correlated in such a way giving vanishing value of $0\nu\beta\beta$ decay amplitude $|M_{ee}|$(Table  (\ref{tab7})). 

\begin{figure}
    \centering
    \includegraphics[width=8cm]{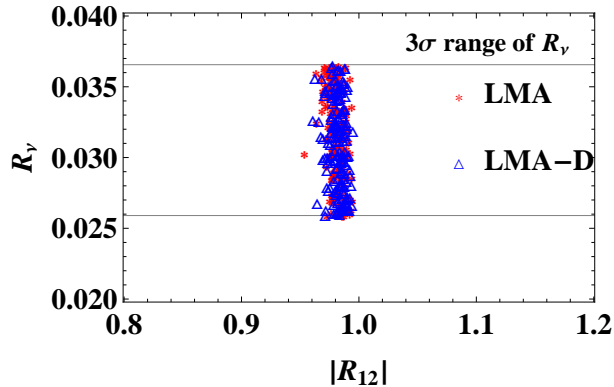}
    \caption{Correlation between ($\left|R_{12}\right|-R_{\nu}$) for $D_2$ texture.}
    \label{fig5}
\end{figure}

\begin{figure}
    \centering
    \includegraphics[width=16cm]{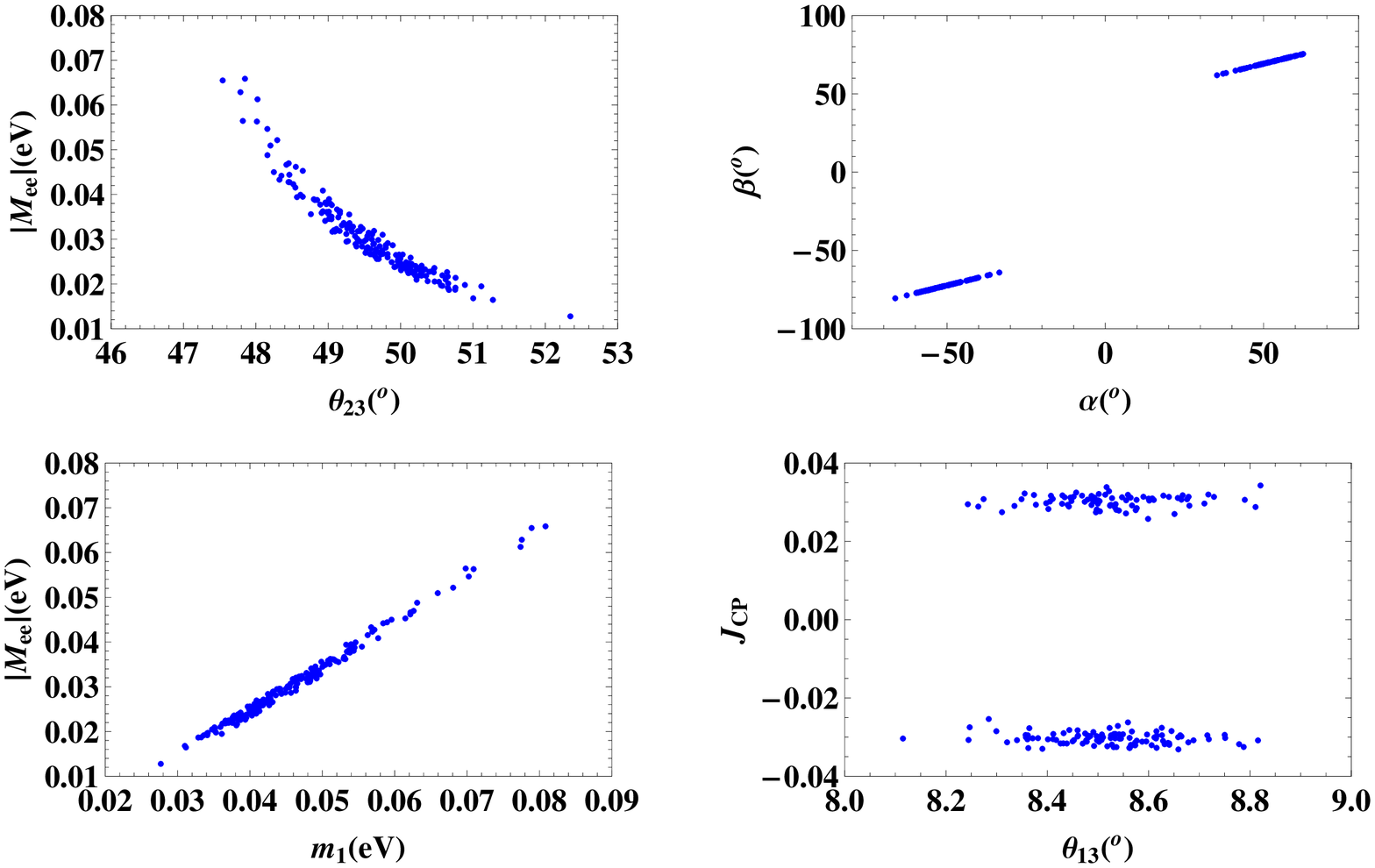}
    \caption{Correlation plots for $D_2$ texture under LMA solution with NH.}
    \label{fig6}
\end{figure}

\begin{figure}
    \centering
    \includegraphics[width=16cm]{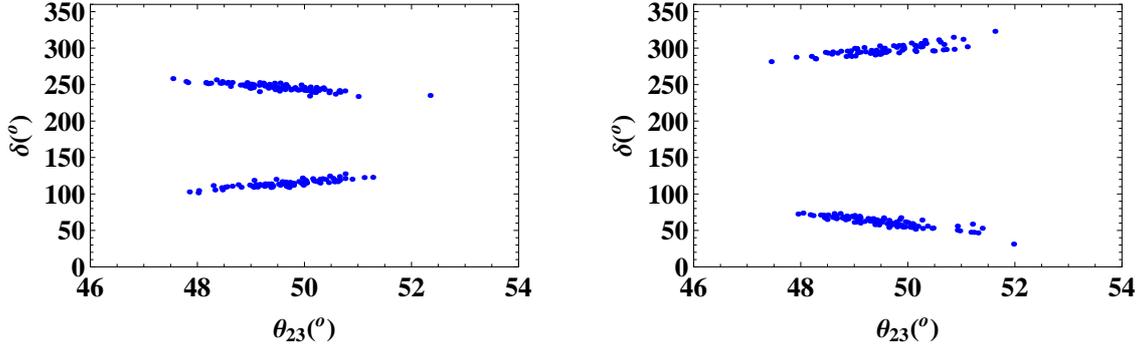}
    \caption{Correlation between $(\theta_{23}-\delta)$ for $D_2$ texture with NH under LMA(left) and LMA-D(right).}
    \label{fig7}
\end{figure}

\noindent Texture $D_2$ is found to be consistent with both LMA and LMA-D descriptions with normal hierarchical neutrino masses, as shown in Figs. (\ref{fig5}) and (\ref{fig6}). It is evident from ($\theta_{23}-|M_{ee}|$) correlation plot in Fig. (\ref{fig6}) that there exist a $3\sigma$ lower bound on $0\nu\beta\beta$ decay amplitude $|M_{ee}|>0.02$ eV (see Table (\ref{tab7})). Furthermore, in contrast to $D_1$, texture $D_2$ is found to be necessarily $CP$ violating as depicted in ($\theta_{13}$-$J_{CP}$) (Fig. (\ref{fig6})) and ($\theta_{23}$-$\delta$) (Fig. (\ref{fig7}))correlation plots.

\subsection{Class E}
The mass ratios for texture $E_1$ can be obtained from Eqns. (\ref{30}) and (\ref{31}) by using the transformation $c_{23}\rightarrow s_{23}; s_{23}\rightarrow c_{23}$ \textit{viz.,}
\begin{eqnarray}
&&|R_{12}|\equiv\dfrac{m_{1}}{m_{2}}\approx \frac{s_{12}^2}{c_{12}^2}+\frac{s_{12}c_{23}s_{13}\cos\delta}{c_{12}^3s_{23}},
\label{32}\\
&&|R_{13}|\equiv\dfrac{m_{1}}{m_{3}}\approx \tan\theta_{12}\tan\theta_{23}s_{13}.
\label{33}
\end{eqnarray}

\begin{figure}
    \centering
    \includegraphics[width=8cm]{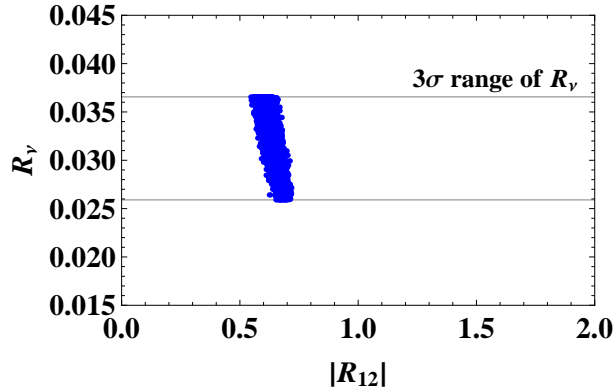}
    \caption{Correlation between ($\left|R_{12}\right|-R_{\nu}$) for texture $E_1$ under LMA solution with NH.}
    \label{fig8}
\end{figure}

\begin{figure}
    \centering
    \includegraphics[width=16cm]{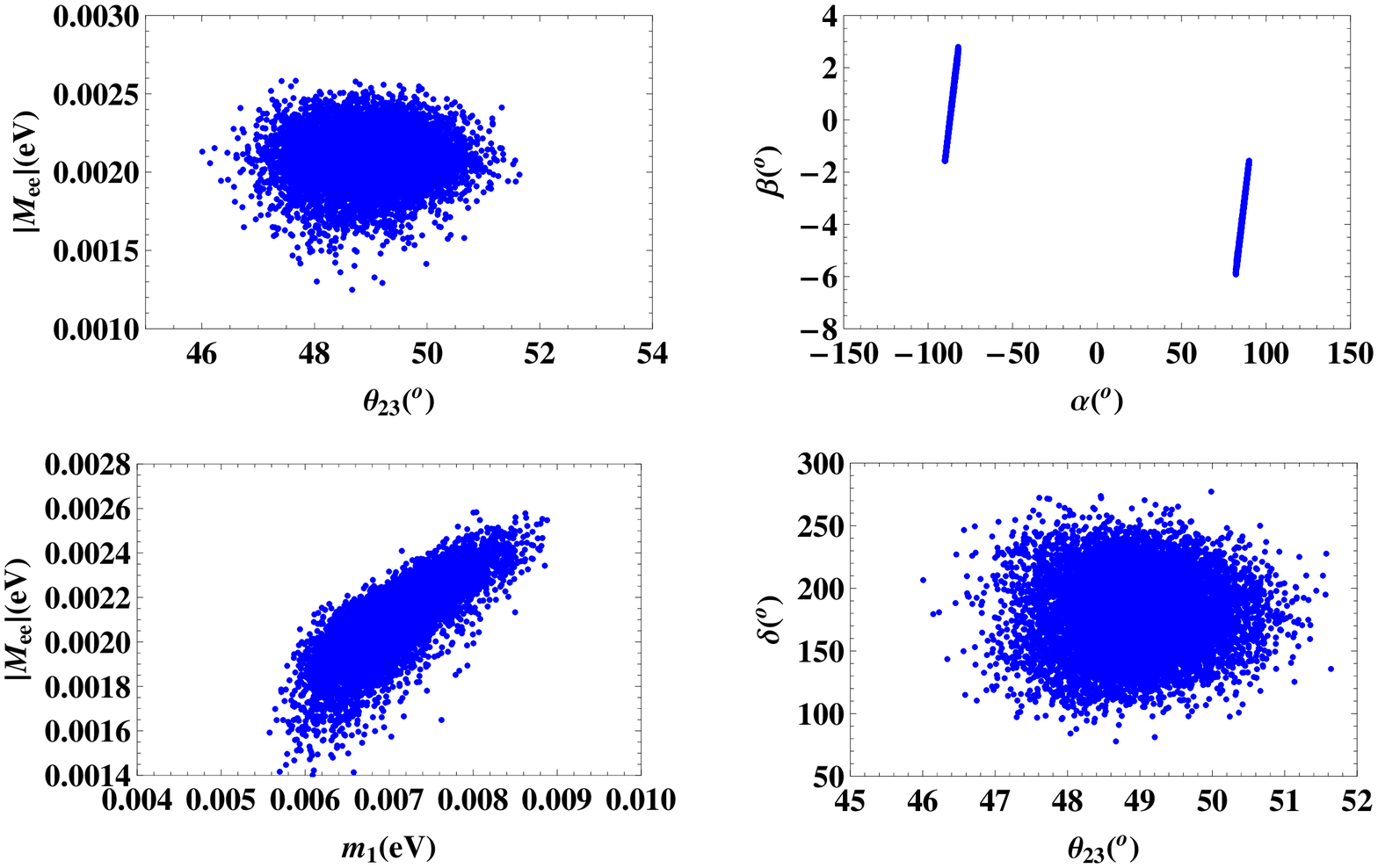}
    \caption{Correlation plots for texture $E_1$ under LMA solution with NH.}
    \label{fig9}
\end{figure}

\noindent It can be seen from Eqn. (\ref{32}) that $E_1$ is not consistent with LMA-D solution as $|R_{12}|>1$. The allowed parameter space for LMA with NH is shown in Figs. (\ref{fig8}) and (\ref{fig9}). Also, the Majorana phases are correlated in such a way that $0\nu\beta\beta$ decay amplitude $|M_{ee}|$ is found to be vanishing in this case. Furthermore, the texture allows for both $CP$ conserving and violating solutions as is evident from Fig. (\ref{fig9}).     

\begin{table}[t]
    \centering
    \begin{tabular}{|c|c|c|c|}
    \hline
Class & Texture & LMA & LMA-D \\
     \hline
   
 \multirow{5}{*}{A} &   $A_1$ (NH/IH)& $\times$/$\times$ & $\times$/$\times$ \\
    \cline{2-4}
 
    &$A_2$ (NH/IH)& $\times$/$\times$ & $\times$/$\times$ \\
   \cline{2-4}
    
   & $A_3$ (NH/IH)& $\times$/$\times$  &$\times$/$\times$ \\
    \cline{2-4}
   
   & $A_4$ (NH/IH)& $\times$/$\times$ &$\times$/$\times$ \\
   \cline{2-4}
   
   & $A_5$ (NH/IH)& $\times$/$\times$ & $\times$/$\times$ \\
   \hline
    
  \multirow{4}{*}{B}&  $B_1$ (NH/IH)& $\times$/$\times$  & $\times$/$\times$\\
   \cline{2-4}
   
   & $B_2$ (NH/IH)&$\checkmark$/$\times$  &$\checkmark$/$\times$ \\
  \cline{2-4}
    
  &  $B_3$ (NH/IH)& $\times$/$\times$  & $\times$/$\times$\\
   \cline{2-4}
    
  &  $B_4$ (NH/IH)&$\times$/$\checkmark$ &$\times$/$\checkmark$ \\
    \hline
   
    \multirow{3}{*}{C}&  $C_1$ (NH/IH)&$\checkmark$/$\times$  &$\checkmark$/$\times$ \\
    \cline{2-4}
   
   & $C_2$ (NH/IH)&$\times$/$\times$  &$\times$/$\times$ \\
     \cline{2-4}
   
   & $C_3$ (NH/IH)&$\times$/$\checkmark$  & $\times$/$\checkmark$\\
    \hline
   
     \multirow{2}{*}{D}& $D_1$ (NH/IH)&$\checkmark$/$\times$  & $\times$/$\times$\\
   \cline{2-4}
  
  &  $D_2$ (NH/IH)& $\checkmark$/$\times$ &$\checkmark$/$\times$ \\
    \hline
  
   E& $E_1$ (NH/IH)&$\checkmark$/$\times$  &$\times$/$\times$ \\
    \hline

    \end{tabular}
    \caption{Allowed/disallowed two-zero textures of $M_{\nu}^{-1}$ under LMA and LMA-D solutions. The $\checkmark$ ($\times$) mark is used to denote allowed (disallowed) texture.}
    \label{tab6}
\end{table}
\begin{table}[t]
\centering
\begin{tabular}{|c|c|c|c|c|}
\hline
\multirow{3}{*}{Allowed Textures}&\multicolumn{4}{c|}{Lower bound on $|M_{ee}|$ (eV)}\\
\cline{2-5}
 & \multicolumn{2}{c|}{LMA} & \multicolumn{2}{c|}{LMA-D} \\
\cline{2-5}
    & NH & IH & NH & IH \\
    \hline
    $B_{2}$ & 0.02 & $\times$ & 0.04 & $\times$ \\
    \hline
    $B_{4}$ & $\times$ & 0.06 & $\times$ & 0.06 \\
    \hline
    $C_{1}$ & 0.02 & $\times$ & 0.02 & $\times$\\
    \hline
    $C_{3}$ & $\times$ & 0.06 & $\times$ & 0.06\\
    \hline
     $D_{1}$ & 0 & $\times$ & $\times$ & $\times$\\
    \hline
     $D_{2}$ & 0.01 & $\times$ & 0.02 & $\times$ \\
    \hline
     $E_{1}$ & 0 & $\times$ & $\times$ & $\times$\\
    \hline
    
  \end{tabular}
  
  \caption{$3\sigma$ lower bound on effective Majorana mass ($|M_{ee}|$) in eV for all allowed textures. \enquote{$\times$} symbolize the disallowed hierarchy in the corresponding texture.}
  \label{tab7}
\end{table}

\newpage
\newpage

\newpage

\newpage
\section{Symmetry Realization}
\noindent In this section, we discuss the minimal realization of two-zero texture of $M_{\nu}^{-1}$. Motivated by the pivotal character played by the discrete flavor symmetry groups in explaining the observed neutrino oscillation data\cite{Ma:2001dn,Ishimori:2012zz}, we obtain the $A_4$ flavor group based symmetry realisation of inverse neutrino mass matrix ($M_{\nu}^{-1}$) by extending the standard model particle content in the lepton sector. $A_4$ is a non-Abelian discrete group of even permutations. $A_4$ group is a orientation-preserving symmetry of a regular tetrahedron. It has four irreducible representations 1, $1'$, $1''$ and 3 and can be generated using S and T generators satisfying the relations
\begin{equation}
    \nonumber
        S^2=T^3=(ST)^3=1.
    \end{equation}
 \noindent Here, we choose basis for $A_{4}$ group in which T generator takes the diagonal form. The reason behind choosing this particular representation is that it facilitates the diagonal mass matrix for charged leptons. In T-diagonal basis, one dimensional unitary representation 1, $1'$ and $1''$ with generator S and T can be written as
\begin{eqnarray}
\nonumber
1:\hspace{0.5cm} S  =  1, \hspace{0.1cm}  T= 1, \end{eqnarray}
\begin{eqnarray}
\nonumber
1':\hspace{0.5cm} S  =  1, \hspace{0.1cm} T=\omega, 
\end{eqnarray}
\begin{eqnarray}
\nonumber
1'':\hspace{0.5cm} S  =  1, \hspace{0.1cm} T=\omega^2, 
\end{eqnarray}
such that $\omega$= $e^{i2\pi/3}$
whereas three-dimensional unitary representation is given by
\begin{equation}
\nonumber
T=
    \begin{pmatrix}
1 &  0 & 0 \\
0 & \omega & 0 \\
0 & 0  & \omega^2\\
\end{pmatrix},\hspace{3mm}
S=\dfrac{1}{3}
    \begin{pmatrix}
-1 & 2 & 2 \\
2 & -1 & 2 \\
2 & 2  & -1\\
\end{pmatrix}.    
\end{equation}
\noindent The multiplication rules for the representations of $A_{4}$ are as follows
\begin{eqnarray}
\nonumber
1'\otimes 1'=1'', \hspace{5mm} 1''\otimes 1''=1' , \hspace{5mm} 1'\otimes 1''=1, \hspace{5mm} 1''\otimes 1=1'',
\end{eqnarray}
\begin{eqnarray}
\nonumber
1\otimes 1'=1', \hspace{5mm} 3\otimes 1'=3 , \hspace{5mm} 3\otimes 1''=3, \hspace{5mm} 3\otimes3=1\oplus 1'\oplus 1''\oplus 3_{s} \oplus 3_{a}.
\end{eqnarray}
In the $T$-diagonal basis, the Clebsch–Gordan decomposition of two triplets, a = ($a_1, a_2, a_3$) and b = ($b_1, b_2, b_3$) is given as
\begin{eqnarray}\label{tp}
\nonumber
&&(a\otimes b)_{1}=a_1b_1+a_2b_3+a_3b_2 ,\\
\nonumber
&&(a\otimes b)_{1'}=a_3b_3+a_1b_2+a_2b_1 ,\\
\nonumber
&&(a\otimes b)_{1''}=a_2b_2+a_1b_3+a_3b_1 ,\\
\nonumber
&&(a\otimes b)_{3_{s}}=\dfrac{1}{3}\left(2a_1b_1-a_2b_3-a_3b_2, 2a_3b_3-a_1b_2-a_2b_1, 2a_2b_2-a_1b_3-a_3b_1\right) ,\\
&&(a\otimes b)_{3_{a}}=\dfrac{1}{2}\left(a_2b_3-a_3b_2, a_1b_2-a_2b_1, a_1b_3-a_3b_1\right).
\end{eqnarray}
\begin{table}[]
    \centering
    \begin{tabular}{|c| c c c c c c c c c c c|c|}
    \hline
    Symmetry & ${D}_{eL}$ &${D}_{\mu L}$&${D}_{\tau L}$& $e_R$ & $\mu_R$ & $\tau_R$ & $\nu_{1 R}$ & $\nu_{2 R}$ & $\nu_{3 R}$ & $\phi$ & $\chi$ &  Obtained two-zero textures \\
     \hline
    $SU(2)_L$ & 2 & 2 & 2 & 1 & 1 & 1 & 1 & 1 & 1 & 2 & 1 &\\
    \hline
    \multirow{2}{*}{$A_4$} & 1 & $1''$ & $1'$ & 1 & $1'$ & $1''$ & 1 & $1'$ & $1''$ & 1 & $1'$ & $B_4$= $\begin{pmatrix}
X &  0 & X \\
0 & X  & X \\
X & X  & 0\\
\end{pmatrix}$  \\
    \cline{2-13}
    & 1 & $1''$ & $1'$ & 1 & $1'$ & $1''$ & 1 & $1'$ & $1''$ & 1 & $1''$ &$C_1$=$\begin{pmatrix}
X & X  & 0 \\
X & 0  & X \\
0 & X  & X\\
\end{pmatrix}$ \\
    \hline
    \end{tabular}
    \caption{Field content and charge assignments in the model under $SU(2)_L$ and $A_4$ symmetries.  }
    \label{tab8}
\end{table}
Here, we have worked in the framework of Type-I seesaw.  In addition, we have minimally extended the standard model by adding three right-handed neutrino fields ($\nu_{iR}$; $i=1,2,3$) and one scalar field ($\chi$), having singlet representation under $A_4$ symmetry as shown in the Table (\ref{tab7}).  In general, for any Yukawa coupling to be non-zero, its Yukawa Lagrangian term must be in singlet-invariant representation of $A_{4}$ with mass dimension four at tree level.\\
\textbf{Texture $B_{4}$:}\\
Using the tensor products in Eqn. (\ref{tp}), the invariant Yukawa Lagrangian is given by
\begin{eqnarray}\label{Lag}
\nonumber
- \mathcal{L}=........&&+y_{e} \bar{D}_{eL} \phi e_{R}+y_{\mu} \bar{D}_{\mu L} \phi \mu_{R}+y_{\tau} \bar{D}_{\tau L} \phi \tau_{R} \\
\nonumber
&&+y_{1} \bar{D}_{eL} \tilde{\phi} \nu_{eR}+y_{2} \bar{D}_{\mu L} \tilde{\phi} \nu_{\mu R}+y_{3} \bar{D}_{\tau L} \tilde{\phi} \nu_{\mu R} \\
\nonumber
&&+\dfrac{1}{2}\left[M_{1}(\nu^{T}_{1R}C^{-1}\nu_{1R})+M_{2}(\nu^{T}_{2 R}C^{-1}\nu_{3R}+\nu^{T}_{3 R}C^{-1}\nu_{2 R})\right]
\\
&&+\dfrac{1}{2}\left[\left(y_{\chi_{1}}(\nu^{T}_{2R}C^{-1}\nu_{2R})+y_{\chi_{2}}(\nu^{T}_{1 R}C^{-1}\nu_{3 R}+\nu^{T}_{3 R}C^{-1}\nu_{1 R})\right)\chi\right],
\end{eqnarray}
where $y_{k}, y_{i}$ ($k=e, \mu, \tau$; $i=1,2,3$) are Yukawa coupling constants, $M_{1,2}$ are bare mass terms for right-handed Majorana neutrinos, $y_{\chi_{1,2}}$ denotes Yukawa coupling constant for interaction terms with scalar field $\chi$ and  $ \tilde{\phi}= i \tau_{2}   \phi^{*}$; $\tau_{2}$ being Pauli matrix.
\noindent The dots in the Lagrangian represents the other kinetic and scalar potential terms. We have restricted up to Yukawa interactions pertaining to mass terms. Spontaneous symmetry breaking (SSB) occurred with vacuum expectation values ($vev$'s) $v$ and $w$ for the Higgs doublet and scalar singlet field, respectively. The Yukawa Lagrangian (Eqn. (\ref{Lag})) leads to the mass matrices as 
\begin{eqnarray} \label{Mlmd}
M_{l}=
\begin{pmatrix}
y_e v &  0 & 0 \\
0 & y_{\mu}v  & 0 \\
0 & 0  & y_{\tau} v\\
\end{pmatrix}, \hspace{1cm}
M_{D}=
\begin{pmatrix}
y_1 v &  0 & 0 \\
0 & y_{2} v  & 0 \\
0 & 0  & y_{3}v\\
\end{pmatrix}
\label{eq25}
\end{eqnarray}
and
\begin{eqnarray}\label{MR}
M_{R}=
\begin{pmatrix}
M_{1} & 0 & y_{\chi_{2}} w \\
0 & y_{\chi_{1}}w  & M_{2} \\
y_{\chi_{2}}w & M_{2}  & 0 \\
\end{pmatrix},
\end{eqnarray}
where $M_{l}$, $M_{D}$ and $M_{R}$ corresponds to charged lepton mass matrix, Dirac mass matrix and right-handed Majorana mass matrix, respectively.\\
\noindent Type-I seesaw contribution to effective Majorana neutrino mass matrix is given by
\begin{eqnarray}
M_{\nu} = M_{D} M_{R}^{-1} M_{D}^{T}.
\end{eqnarray}

\noindent Also, the inverse neutrino mass matrix can be written as
\begin{eqnarray}
M_{\nu}^{-1}=M_{D}^{-T}M_{R}M_{D}^{-1}.
\end{eqnarray}
\noindent In $M_{D}$-diagonal basis, the peculiar feature of implementation of type-I seesaw for $M_{\nu}^{-1}$ is that the zero(s) in $M_{R}$ corresponds to zero(s) in $M_{\nu}^{-1}$.
Using the Eqns. (\ref{Mlmd}) and (\ref{MR}), the $M_{\nu}^{-1}$ is given by
\begin{eqnarray}
M^{-1}_{\nu}=
\begin{pmatrix}
\dfrac{M_{1}}{v^2 y^{2}_{1}} & 0 & \dfrac{y_{\chi_{1}}w}{v^2 y_{1} y_{3}} \\
0 & \dfrac{y_{\chi_{2}}w}{v^{2} y^{2}_{3}}  & \dfrac{M_{2}}{v^{2}y_{2}y_{3}} \\
\dfrac{y_{\chi_{1}}w}{v^{2} y_{1}y_{3}} & \dfrac{M_{2}}{v^{2}y_{2}y_{3}} & 0 \\
\end{pmatrix},
\end{eqnarray}
which symbolically can be written as
\begin{eqnarray}
M_{\nu}^{-1}=
\begin{pmatrix}
X & 0  & X  \\
0 & X & X  \\
X & X & 0
\end{pmatrix}
\end{eqnarray}
 corresponding to texture $B_{4}$.\\
\textbf{Texture $C_{1}$:}\\
For the realization of texture $C_1$, we change the irreducible representation of scalar field $\chi$ to be $1''$.
The relevant Yukawa Lagrangian is  
\begin{eqnarray}
\nonumber
- \mathcal{L}=........&&+y_{e} \bar{D}_{eL} \phi e_{R}+y_{\mu} \bar{D}_{\mu L} \phi \mu_{R}+y_{\tau} \bar{D}_{\tau L} \phi \tau_{R} \\
\nonumber
&&+y_{1} \bar{D}_{eL} \tilde{\phi} \nu_{eR}+y_{2} \bar{D}_{\mu L} \tilde{\phi} \nu_{\mu R}+y_{3} \bar{D}_{\tau L} \tilde{\phi} \nu_{\mu R} \\
\nonumber
&&+\dfrac{1}{2}\left[M_{1}(\nu^{T}_{1R}C^{-1}\nu_{1R})+M_{2}(\nu^{T}_{2 R}C^{-1}\nu_{3R}+\nu^{T}_{3 R}C^{-1}\nu_{2 R})\right]
\\
&&+\dfrac{1}{2}\left[\left(y_{\chi_{1}}(\nu^{T}_{3R}C^{-1}\nu_{3R})+y_{\chi_{2}}(\nu^{T}_{1 R}C^{-1}\nu_{2 R}+\nu^{T}_{2 R}C^{-1}\nu_{1 R})\right)\chi\right]
\end{eqnarray}
where $y_{k}, y_{i}$($k=e, \mu, \tau$; $i=1,2,3$) are Yukawa coupling constants, $M_{1,2}$ are bare mass terms for right-handed Majorana neutrinos, $y_{\chi_{1,2}}$ denotes Yukawa coupling constant for interaction terms with scalar field $\chi$ and  $ \tilde{\phi}= i \tau_{2}   \phi^{*}$; $\tau_{2}$ being Pauli matrix.\\
After SSB, charged lepton mass matrix and Dirac mass matrix remains diagonal as shown in Eqn.  (\ref{eq25}). But Majorana mass matrix gets modified and takes the form
\begin{eqnarray}
M_{R}=
\begin{pmatrix}
M_{1} & y_{\chi_{2}}w & 0 \\
y_{\chi_{2}}w & 0  & M_{2} \\
0 & M_{2}  & y_{\chi_{1}}w \\
\end{pmatrix}.
\end{eqnarray}
In $M_{D}$-diagonal basis, the inverse neutrino mass matrix is given by
\begin{eqnarray}
M^{-1}_{\nu}=
\begin{pmatrix}
\dfrac{M_{1}}{v^2 y^{2}_{1}} & \dfrac{y_{\chi_{2}}w}{v^2 y_{1} y_{2}} & 0 \\
\dfrac{y_{\chi_{2}}w}{v^2 y_{1} y_{2}} & 0  & \dfrac{M_{2}}{v^{2}y_{2}y_{3}} \\
0 & \dfrac{M_{2}}{v^{2}y_{2}y_{3}} & \dfrac{y_{\chi_{1}}w}{v^{2} y^{2}_{3}} \\
\end{pmatrix},
\end{eqnarray}
which symbolically can be written as
\begin{eqnarray}
M^{-1}_{\nu}=
\begin{pmatrix}
X & X  & 0 \\
X & 0  & X \\
0 & X  & X\\
\end{pmatrix},
\end{eqnarray}
corresponding to texture $C_{1}$.
\section{Conclusions}

In conclusion, we have investigated the phenomenological implications of two-zero textures  in inverse neutrino mass matrix $M_{\nu}^{-1}$. In the basis where Dirac neutrino mass matrix $M_D$ is diagonal, zeros in right-handed Majorana neutrino mass matrix $M_R$ corresponds to zeros of $M_{\nu}^{-1}$. We have, also, proposed symmetry realization, based on discrete flavor group $A_4$, wherein such texture zeros  can emerge in $M_{\nu}^{-1}$. Further, we investigate the viability of all two-zero textures in $M_{\nu}^{-1}$ under LMA and LMA-D solutions. We have categorized the possible textures in five classes \textit{viz.} class A, B, C, D and E. Out of fifteen possible two-zero textures of $M_{\nu}^{-1}$, seven are found to be in consonance with LMA and/or LMA-D scenario.  The general remarks about the obtained phenomenology are as under:  
\begin{itemize}
    \item The textures in class A are all disallowed as they do not reproduce the correct neutrino phenomenology. Thus, texture with $\left(M_{\nu}^{-1}\right)_{11}=0$ is disallowed, in general.
    \item In class B, $B_2$ and $B_4$ textures are consistent with both LMA and LMA-D solutions. $B_{2}$ ($B_4$) predicts normal (inverted) hierarchical neutrino masses. For texture $B_2$, the $3\sigma$ lower bound on $0\nu\beta\beta$ decay amplitude $|M_{ee}|$ is found to be $0.02$ eV ($0.04$ eV) under LMA and LMA-D, respectively. For $B_4$ texture, it is about $0.06$ eV for both LMA and LMA-D solutions. 
    \item In class C, $C_2$ is disallowed. $C_1$ and $C_3$ textures are consistent with both LMA and LMA-D solutions. $C_1$ ($C_3$) predicts normal (inverted) hierarchical neutrino masses. For texture $C_1$ ($C_3$), the $3\sigma$ lower bound on $|M_{ee}|$ is $0.02$ eV ($0.06$) eV under both LMA and LMA-D solutions.
    \item For textures $B_2$, $B_4$, $C_1$ and $C_3$, the Dirac and Majorana-type $CP$ violating phases are sharply constrained and these textures are found to be necessarily $CP$ violating.
    \item Textures $D_1$ and $E_1$ predict normal hierarchical neutrino masses and are found to be consistent with LMA solution. LMA-D solution is disallowed by these textures. Also,  $|M_{ee}|$ is vanishing in these textures. In general, we conclude that the textures for which LMA-D is disallowed, $|M_{ee}|$ is vanishing. Similar inference is observed in Ref. \cite{Borgohain:2020now} wherein the authors analyzed phenomenology of Majorana neutrino textures in the light of LMA-D solution.
    \item Texture $D_2$ predicts normal hierarchical neutrino masses and is consistent with both LMA and LMA-D phenomenology. In LMA (LMA-D) scenario, there exist a $3\sigma$ lower bound on $|M_{ee}|>0.01$ eV ($0.02$ eV).
    \item The generic feature of the class of model, discussed in the present work, is the existence of neutrino mass hierarchy degeneracy in a particular texture. For example, if a texture is allowed by LMA solution with ``X" neutrino mass hierarchy then, if LMA-D is allowed, it is allowed with the same hierarchy ``X".

\end{itemize}
The allowed two-zero texture of $M_{\nu}^{-1}$ \textit{viz.} $B_2$, $B_4$, $C_1$, $C_3$, $D_1$, $D_2$ and $E_1$ has imperative predictions for $|M_{ee}|$\cite{onbb}. Except for $D_1$ and $E_1$ textures, the predicted $3\sigma$ lower bound on $0\nu\beta\beta$ decay amplitude $|M_{ee}|$ is $\mathcal{O}(10^{-2})$ which is within the sensitivity reach of $0\nu\beta\beta$ decay experiments like SuperNEMO\cite{snamo}, KamLAND-Zen\cite{kzen}, NEXT\cite{nxt1,nxt2}, and nEXO\cite{nexo}. For example, the non-observation of $0\nu\beta\beta$ decay down to these high sensitivities will refute all the textures except $D_1$ and $E_1$. Also, we have shown that the allowed $M_{\nu}^{-1}$ textures can be accommodated in an extension of the SM with three right-handed neutrinos and one scalar singlet field. As representative realizations, we have obtained two such textures $B_{4}$ and $C_{1}$ within Type-I seesaw scenario using $A_{4}$ discrete flavor symmetry.

\noindent\textbf{\Large{Acknowledgments}}
 \vspace{.3cm}\\
 M. K. acknowledges the financial support provided by Department of Science and Technology(DST), Government of India vide Grant No. DST/INSPIRE Fellowship/2018/IF180327. The authors, also, acknowledge Department of Physics and Astronomical Science for providing necessary facility to carry out this work.


\end{document}